\documentclass[aps,nofootinbib,prd,eqsecnum,showpacs,showkeys,preprintnumbers]{revtex4-1}
\usepackage{graphicx}
\usepackage{amsmath}
\usepackage{amsfonts}
\usepackage{amssymb}
\usepackage{color}
\usepackage{bm}
\usepackage{float}
\usepackage{mathrsfs}
\usepackage{epstopdf}
\usepackage{url}
\usepackage{booktabs}
\usepackage{footnote}
\usepackage{textcomp}
\usepackage[normalem]{ulem}
\usepackage{esint}
\usepackage[unicode=true, pdfusetitle,
 bookmarks=true,bookmarksnumbered=false,bookmarksopen=false,
 breaklinks=false,pdfborder={0 0 1},backref=false,colorlinks=false]{hyperref}
\usepackage{multirow}
\usepackage{pifont}
\usepackage{times}
\usepackage[english]{babel}

\usepackage{appendix}
\usepackage{float}
\usepackage{enumerate}
\usepackage{lineno}

\usepackage{hyperref}

\usepackage{tabularx}

\makeatletter

\newcommand{\stkout}[1]{\ifmmode\text{\sout{\ensuremath{#1}}}\else\sout{#1}\fi}

\AtBeginDocument{
\heavyrulewidth=.08em
\lightrulewidth=.05em
\cmidrulewidth=.03em
\belowrulesep=.65ex
\belowbottomsep=0pt
\aboverulesep=.4ex
\abovetopsep=0pt
\cmidrulesep=\doublerulesep
\cmidrulekern=.5em
\defaultaddspace=.5em
}

\newcolumntype{L}[1]{>{\hsize=#1\hsize\raggedright\arraybackslash}X}%
\newcolumntype{R}[1]{>{\hsize=#1\hsize\raggedleft\arraybackslash}X}%
\newcolumntype{C}[1]{>{\hsize=#1\hsize\centering\arraybackslash}X}%

\newcommand*\patchAmsMathEnvironmentForLineno[1]{%
 \expandafter\let\csname old#1\expandafter\endcsname\csname #1\endcsname
 \expandafter\let\csname oldend#1\expandafter\endcsname\csname end#1\endcsname
 \renewenvironment{#1}%
   {\linenomath\csname old#1\endcsname}%
   {\csname oldend#1\endcsname\endlinenomath}}%
\newcommand*\patchBothAmsMathEnvironmentsForLineno[1]{%
 \patchAmsMathEnvironmentForLineno{#1}%
 \patchAmsMathEnvironmentForLineno{#1*}}%
\AtBeginDocument{%
\patchBothAmsMathEnvironmentsForLineno{equation}%
\patchBothAmsMathEnvironmentsForLineno{align}%
\patchBothAmsMathEnvironmentsForLineno{flalign}%
\patchBothAmsMathEnvironmentsForLineno{alignat}%
\patchBothAmsMathEnvironmentsForLineno{gather}%
\patchBothAmsMathEnvironmentsForLineno{multline}%
}

\begin{document}

\title{Cosmological constraints of phantom dark energy models}

\author{Amine Bouali$^{1}$}
\email{aminebouali\_smp@yahoo.com}
\author{Imanol Albarran $^{2,3}$}
\email{imanolalbarran@gmail.com}
\author{Mariam Bouhmadi-L\'{o}pez $^{4,5}$}
\email{{\mbox{mariam.bouhmadi@ehu.eus}}}
\author{Taoufik Ouali$^{1}$}
\email{ouali1962@gmail.com}

\date{\today }
\affiliation{
${}^1$Laboratory of Physics of Matter and Radiation, Mohammed I University, BP 717, Oujda, Morocco\\
${}^2$Departamento de F\'{\i}sica, Universidade da Beira Interior, Rua Marqu\^{e}s D'\'Avila e Bolama 6200-001 Covilh\~a, Portugal\\
${}^3$Centro de Matem\'atica e Aplica\c{c}\~oes da Universidade da Beira Interior, Rua Marqu\^{e}s D'\'Avila e Bolama 6200-001 Covilh\~a, Portugal\\
${}^4$Department of Theoretical Physics University of the Basque Country UPV/EHU. P.O. Box 644, 48080 Bilbao, Spain\\
${}^5$IKERBASQUE, Basque Foundation for Science, 48011, Bilbao, Spain\\
}

\begin{abstract}
We address three genuine phantom dark energy models where each of them induces the particular future events known as Big Rip, Little Rip and Little Sibling of the Big Rip. The background models are fully determined by a given dark energy equation of state. We first observationally constrain  the corresponding model  parameters that characterise each paradigm  using the available data of supernova type Ia, Cosmic Microwave Background and Baryonic Acoustic Oscillations by using a Markov Chain Monte Carlo method. The obtained fits are used to solve numerically the first order cosmological perturbations. We compute the evolution of the density contrast of  (dark) matter and DE, from the radiation dominated era till a totally DE dominated universe. Then, the obtained results are compared with respect to $\Lambda$CDM.  We obtain the predicted current matter power spectrum and the evolution of $f\sigma_8$ given by the models studied in this work. Finally, the models are tested by computing the reduced $\chi^2$  for the  ``Gold2017'' $f\sigma_8$ dataset.
\end{abstract}

\maketitle

\section{introduction}
It is well known that the expansion of our Universe is accelerating. This fact was observationally supported firstly in 1998 by the measurements of supernova type Ia (SNIa) \cite{Riess:1998cb,Perlmutter:1998np} and then, corroborated by  measurements of the Cosmic Microwave Background (CMB)  and Baryonic Acoustic Oscillations (BAO)  \cite{Ade:2015rim}. On the other hand, the history of the expansion reveals that the transition to an accelerated state happened recently \cite{Ade:2015xua,Aghanim:2018eyx}. The origin of the matter that induces the current speed up of the Universe is still unknown and it is usually dubbed dark energy (DE) \cite{Tsujikawa:2010sc,AmendolaTsujikawa}. In addition, the contribution of  DE to the total energy density of the Universe is roughly  $70\%$ \cite{Ade:2015rim,Ade:2015xua}, so the hidden source that causes the current acceleration of the Universe covers a significant portion of the total energy budget. We do  not know much about the fundamental  cause of dark energy. However, there are phenomenological models that can describe suitably the current expansion of the Universe. Amazingly, the cosmological constant suggested by Einstein, in principle, to get a static Universe, becomes nowadays the paradigm that  better fits the observational data. The model, which also takes into account the contribution of dark matter (DM) was coined as $\Lambda$CDM. This model is characterised by having an Equation of State (EoS) parameter for dark energy which is  constant and equal to $-1$, in such a way that the asymptotic evolution leads to a de Sitter Universe. Despite that the $\Lambda$CDM model gives the best observational fit, there is no reason to exclude other models that could describe as well suitably the current acceleration. In addition, the $\Lambda$CDM model suffers from some fine tuning problems.

It can be said that the problem  has become the problem of the century for cosmologist. The urge to reveal this intriguing fact has motivated a vast amount of works trying to explain the recent speed up of the Universe. In this way, several models that can induce a positive acceleration have been suggested.  We can classify  them in two groups: (i) quintessence models  which are those that preserve the null energy condition, i.e. $0\leq\rho+p$, in such a way that the EoS parameter is always larger than $-1$. (ii)  phantom models  where the null energy condition is violated and the EoS parameter can go below $-1$ \cite{Stefancic:2003rc,Dabrowski:2003jm,Elizalde:2008yf}. Surprisingly, phantom models are not excluded, but even seem to be favoured by observations  \cite{Caldwell:1999ew,Jimenez:2016sgs,Aghanim:2018eyx,Sahni:2014ooa,Vagnozzi:2018jhn,Alam:2016wpf}. 

On the other hand, the discovery of an accelerated Universe has opened the door to theorise about an infinity of possible doomsdays, where the universal acceleration  is so powerful that the Universe ends ripping itself in a dramatic state. Those events are named and classified as \textit{singularities} or \textit{abrupt events} depending if they occur at a finite or an infinite cosmic time, respectively. In particular, we will focus on three genuine phantom models, where each of them induce a specific doomsday known as Big Rip (BR) \cite{Caldwell:1999ew,Dabrowski:2003jm,Starobinsky:1999yw,Caldwell:2003vq,
Carroll:2003st,Chimento:2003qy,GonzalezDiaz:2003rf,GonzalezDiaz:2004vq,Sahni:2002dx}, 
Little Rip (LR) \cite{Ruzmaikina,Bouhmadi-Lopez:2013nma,Nojiri:2005sx,Nojiri:2005sr,Stefancic:2004kb,BouhmadiLopez:2005gk,Frampton:2011sp,Brevik:2011mm,Contreras:2018two} and Little Sibling of the Big Rip (LSBR) \cite{Bouhmadi-Lopez:2014cca,Morais:2016bev,Bouhmadi-Lopez:2018lly}.  While a BR is a true singularity, we highlight that a  LR and  a LSBR are abrupt events. However, in all these models the bound structures will be ripped apart unavoidably sooner or later. In this scenario, the energy density  could increase up to the Planck scale, where  quantum effects are expected to be important. This has lead to carry a quantum analysis close to the cosmological singularities/abrupt events, where the classical singularity could be avoided in the quantum  realm \cite{Dabrowski:2006dd,Kamenshchik:2007zj,BouhmadiLopez:2009pu,Kamenshchik:2012ij,Kamenshchik:2013naa,Bouhmadi-Lopez:2013tua,Albarran:2015tga,Albarran:2015cda,Albarran:2016ewi,Bouhmadi-Lopez:2016dcf,Albarran:2017swy,Alonso-Serrano:2018zpi,Bouhmadi-Lopez:2018tel,Albarran:2018mpg,Elizalde:2004mq,Elizalde:2005ju} (see the recent review \cite{Bouhmadi-Lopez:2019zvz}).  In this work, we allude to the models that induce these events as model A, B and C, respectively. In particular, the model A is actually the model known as $w$CDM, where its EoS parameter is constant and less than $-1$. The corresponding model parameters were observationally constrained in  \cite{Ade:2015xua,wikiesa,Aghanim:2018eyx}. On the other hand, the model B was constrained in  \cite{Frampton:2011sp}, where the authors fit observationally the model parameters and compute when different bound  structures are destroyed. However, the model C has not been observationally constrained so far.
All these models need to be compared and fitted with the available observational data. In addition, the  cosmological perturbations have been a useful tool for cosmologist in this task, for example, they predict the matter distribution that can be compared with the observations. The predicted observables within the cosmological  perturbations theory have been widely used to test several models of DE, as well as DE-DM interacting models and $f(R)$ modified gravity.

In the models studied in \cite{Balcerzak:2012ae,Denkiewicz:2014aca,Denkiewicz:2015nai,Denkiewicz:2017ixd}, it is assumed a dependence of the scale factor with cosmic time. In \cite{Balcerzak:2012ae} the authors disregard  DE perturbations and the predicted evolution of the growth rate is compared with observations. In \cite{Denkiewicz:2014aca,Denkiewicz:2015nai}, DE and DM perturbations are considered.  These models are suitable to describe the Universe from the matter dominated epoch till the present time. In addition, most of them predict future singularities or abrupt events. In \cite{Astashenok:2012iy} the authors compute the matter and DE perturbations and fit the results with the observational data. In \cite{Kunz:2006wc}  a mixture of phantom and standard fluids is studied in order to analyse the instabilities arisen at the perturbative level. Some instabilities merge when dealing with a negative EoS parameter of DE fluids. To avoid such instabilities,  in \cite{Bean:2003fb,Valiviita:2008iv} the authors suggest a  decomposition of the pressure in its adiabatic and non adiabatic contributions. In  \cite{Albarran:2016mdu} this method is applied and initial conditions are imposed at the radiation dominated epoch. On the other hand, in \cite{Velten:2017mtr} the authors analyse the effects of non adiabaticity on the growth rate for several DE models and compute the observable $f\sigma_{8}$.

There are  other interesting  models of DE that have been studied within the framework of cosmological perturbations. In \cite{Maeder:2018xww} the authors obtain the growth rate in the framework of a scale invariant theory. The initial conditions are set at a matter dominated epoch and they compute the resulting perturbations for a range of different backgrounds. The DE-DM interacting models are useful to describe a transition to an accelerated Universe \cite{Mifsud:2017fsy,Dutta:2017wfd,Ferreira:2018wup}. In \cite{Mifsud:2017fsy} the authors focus on computing the anisotropies of the CMB and find a particular footprints of the model studied. In \cite{Dutta:2017wfd}, models arising from the scalar-fluid theories with a derivative coupling  are analysed. The authors compute the perturbations and predict particular footprints on the growth rate. On the other hand, in \cite{Ferreira:2018wup} the authors study the perturbations for a model where a DM superfluid is assumed to be responsible of the current acceleration. Such superfluid consists on a combination of the ground and excited states of DM. The obtained expansion history and growth rate are compared with that given by $\Lambda$CDM.

A large class of modified gravity models have been studied. For example, in \cite{Battye:2017ysh,Morais:2015ooa} the authors consider different $f(R)$ models with a non vanishing  anisotropic stress tensor. The impact of the EoS parameter in several perturbation variables is studied and the predicted anisotropies on CMB are faced against observations. In \cite{Arjona:2018jhh} the authors not only constrain observationally the background model but also compute the resulting perturbation variables such as the Bardeen potentials and $f\sigma_{8}$.

On the other hand, the cosmological perturbations are useful to constrain further observationally DE models. For example, in \cite{Denkiewicz:2017ixd} the scale dependent DE perturbations are studied for different DE models where some future singularities are involved. The authors find the possibility to distinguish different DE models in the oncoming missions as DESI, \textit{Euclid}, and \textit{WFirst-2.4}. In \cite{delaMacorra:2018zbk} the authors constrain observationally  a DE scalar field representation in the so called bound DE model.

The most considered observational probes of dark energy are SNIa, BAO and CMB. SNIa describe the expansion history of the universe by means of luminosity distances. BAO have been also developed and provides a direct measurement of the Hubble expansion, $H(z)$, and the angular diameter distance. CMB provides distance priors which are a strong constraint on DE. In order to avoid degeneracy in the observational data, a tighter constraints on the model parameters are obtained by combining all of SNIa, CMB, BAO and measurement of $H(z)$, i.e. the Hubble expansion. In addition, since the observational data are obtained from independent cosmological probes, their total likelihood is the product of each separate likelihoods.

%
In this work, we focus in two goals. The first one consists on  constraining observationally model parameters that are  characterised by models A, B and C using, for consistency, the same samples of data. Indeed, We compare and classify these models with respect to the $\Lambda$CDM and test their consistency  to the  observational data. In order to obtain the best fit parameters, their mean values and their uncertainties, we confront our DE models by means of a Markov Chain Monte Carlo (MCMC) \cite{Arjona:2018jhh} method to the observational data. We use the Pantheon compilation of SNIa dataset \cite{Scolnic:2017caz}, the Planck 2018 distance priors of CMB \cite{Zhai:2018vmm,Aghanim:2018eyx}, the BAO data\footnote{The authors of \cite{Kazantzidis:2018jtb} have regrouped in a chronological order 30 BAO non correlated data points.} including (6dFGS+SDSS+BOSS-LOWZ+BOSS-CMASS+WiggleZ+BOSS-DR12) \cite{Anderson:2013zyy,Beutler:2011hx,Ross:2014qpa,Kazin:2014qga,Alam:2016hwk} and measurements of the Hubble rate \cite{Anderson:2013zyy,Zhang:2012mp,Stern:2009ep,Moresco:2012jh,Chuang:2012qt,Moresco:2015cya,Moresco:2016mzx,Stocker:2018avm}. For the second goal, we will use the previous  best fit parameters to compute the first order linear perturbations and analyse the distribution of matter during the different cosmological epochs. The aim is to analyse the footprints that these models could leave on the distribution of galaxies. Indeed, we compute the predicted matter power spectrum and the evolution of $f\sigma_{8}$ quantity at low redshift. This $f\sigma_{8}$ evolution is faced against a second and independent set of observations (matter power spectrum and $f\sigma_{8}$ data set). For consistency, aside from the models A, B and C, we compute as well the results for the $\Lambda$CDM setup, which we use as a pattern to compare with the other three models.

%
The paper is organised as follows: In section ~\ref{datadescription}, we discuss the details of the different data. In section~\ref{DEmodels}, we introduce the background of the different models under consideration. In section~\ref{results}, we show the obtained results for the model parameters with their confidence levels and contourplots. In section~\ref{pertresults}, we compute the evolution of the perturbation variables and the predicted matter power spectrum and $f\sigma_{8}$. In section~\ref{conclusions}, we discuss our main conclusions. Finally, in the Appendix~\ref{seclinearpert}, we briefly introduce the linear cosmological perturbation theory together with the used initial conditions for the perturbations.

\section{Data description}\label{datadescription}
\subsection{SNIa data}

We have used the Pantheon compilation as a SNIa dataset, they are made of 1048 spectroscopically confirmed Type Ia Supernovae distributed in the redshift interval $0.01<z<2.26$ \cite{Scolnic:2017caz}. The Pantheon sample is the largest compilation up to date and consists of a different supernovae surveys, including SDSS, SNLS, various low-$z$ samples and some high-$z$ samples from HST. 
The distance modulus for supernovae is given by, 
\begin{equation}
\mu_{th}=5\log_{10} \frac{d_L}{Mpc} +25,
\end{equation}
where $d_L=\left( c/H_0 \right) D_L$ is the luminosity distance, $H_0$ is the Hubble constant, $c$ is the speed of light,
\begin{equation}
D_L=(1+z_{\textrm{hel}})\int_{0}^{z_{\textrm{CMB}}}\frac{dz}{E(z)},
\end{equation}
where $z_{\textrm{hel}}$ is the heliocentric redshift and $z_{\textrm{CMB}}$ is the CMB frame redshift, $E(z)=H(z)/H_0$ is the normalized Friedmann equation which encodes DE models.\\

The observed apparent magnitude for the Pantheon compilation is given by $m_{obs}=\mu_{obs}+M$ \cite{Scolnic:2017caz}, where $\mu_{obs}$ is the observed distance modulus and $M$ is the absolute magnitude.
To estimate the cosmological parameters, we compute the  \textit{chi-square}, $\chi^2$.  For SNIa, $\chi_{SN}^2$ is constructed as 
\begin{equation}
\bf \chi_{SN}^2=(\mu_{obs}-\mu_{th})^{T}.C_{Pantheon}^{-1}.(\mu_{obs}-\mu_{th}),
\end{equation}
where $(\mu_{obs}-\mu_{th})$ is the difference vector between the model expectations and the observed magnitudes, where $\bf C_{Pantheon}$ is the covariance matrix of Pantheon data which  is given by the sum of a statistical part and systematic part  $\bf C_{Pantheon}=\bf C_{stat}+\bf C_{sys}$. In order to get rid of the nuisance parameter $M$, we perform an analytical marginalization over it, by defining a new chi-square \cite{Conley:2011ku}
 \begin{equation}
{\bf  \chi_{SN}^2}={\bf{A}}+\ln {\frac{{\bf{C}}}{2\pi}}-\frac{\bf B^2}{ \bf C},
 \end{equation}
 where $\bf A=(\mu_{obs}-\mu_{th})^{T}.C_{Pantheon}^{-1}.(\mu_{obs}-\mu_{th})$,   $\bf B=(\mu_{obs}-\mu_{th})^{T}.C_{Pantheon}^{-1}. 1$ and    $\bf C=1^{T}.C_{Pantheon}^{-1}.1\hspace{0.2cm}$ being $\bf 1$ the 1048$\times$1048 identity matrix.\\ 

\subsection{CMB data}  

The power spectrum of CMB affects crucially the physics, from the decoupling epoch till today. These effects are mainly quantified by the acoustic scale $l_a$ and the shift parameter $R$ given by \cite{Komatsu:2008hk}
\begin{equation}
R\equiv\sqrt{\Omega_{\textrm{m}} H_0^{2}}(1+z_{\textrm{CMB}})D_{A}(z_{\textrm{CMB}}),
\end{equation} 
\begin{equation}
l_a\equiv(1+z_{\textrm{CMB}})\frac{\pi D_{A}(z_{\textrm{CMB}})}{r_s(z_{\textrm{CMB}})}.
\end{equation}
where $z_{\textrm{CMB}}$ is the redshift at the decoupling epoch, $D_{A}(z_{\textrm{CMB}})$ is
the angular diameter distance of photons  in a flat Friedmann-Lema\^itre-Robertson-Walker (FLRW) universe  expressed as
\begin{equation}
D_A(z)=\frac{1}{H_0(1+z)}\int_{0}^{z}\frac{dz'}{E(z')},
\end{equation}
 and $r_s(z)$ is the comoving sound horizon given by 
\begin{equation}\label{rs}
r_s(z)=\frac{1}{H_0}\int_{0}^{a}\frac{da'}{a'E(a')\sqrt{3(1+\bar{R_{b}})a'}},
\end{equation}
where $a=(1+z)^{-1}$ is the conversion rule from the redshift to the scale factor and $R_{b}=31500\Omega_{\textrm{b}} h^2(T_{\textrm{CMB}}/2.7 K)^{-4}$, with $T_{\textrm{CMB}}=2.275 K$ \cite{Fixsen:2009ug}. The redshift at decoupling is given by the fitting formula \cite{Hu:1995en}
\begin{equation}
z_{\textrm{CMB}}=1048[1+0.00124(\Omega_{\textrm{b}} h^2)^{-0.738}][1+g_1(\Omega_{\textrm{m}} h^2)^{g_2}],
\end{equation}
where
\begin{equation}
g_{1}=\frac{0.0783(\Omega_{\textrm{b}} h^2)^{-0.238}}{1+39.5(\Omega_{\textrm{b}} h^2)^{0.763}},
\end{equation}
and
\begin{equation}
g_{2}=\frac{0.56}{1+21.1(\Omega_{\textrm{b}} h^2)^{1.81}}.
\end{equation}

The CMB covariance matrix is given by \cite{Zhai:2018vmm}
\begin{equation}
\bf C_{\textrm{CMB}}=10^{-8}\times\left( 
\begin{array}{ccc}
1598.9554 & 17112.007 & -36.311179 \\ 
17112.007 & 811208.45 & -494.79813 \\ 
-36.311179 &-494.79813 & 2.1242182
\end{array} \right). 
\end{equation}

Finally,  the CMB contribution to the total $\chi^2$ is 
\begin{equation}
\bf \chi_{\textrm{CMB}}^{2}=X_{\textrm{CMB}}^{T}.C_{\textrm{CMB}}^{-1}.X_{\textrm{CMB}},
\end{equation}
where $X_{\textrm{CMB}}$ is the CMB parameters vector based on Planck 2018 release, as derived by \cite{Zhai:2018vmm} 
\begin{equation}
\bf X_{\textrm{CMB}}=\left(
\begin{array}{c}
R-1.74963 \\
l_a-301.80845\\
\Omega_{\textrm{b}} h^2-0.02237
\end{array} 
\right).
\end{equation}

\subsection{BAO data}

The baryon acoustic oscillation  is an important observational data currently used to constrain the cosmological parameters more efficiently in combination with other probes such as the CMB. The information taken from the BAO peaks present in the matter power spectrum can be used to determine the Hubble parameter $H(z)$ and the angular diameter distance $D_A(z)$ which allows us to calculate  DE parameters. The combination of the angular-diameter distance and the Hubble parameter, $D_V(z)$, is given by \cite{Eisenstein:2005su}
\begin{equation}
D_V(z)\equiv\left[(1+z)^2 D_A^{2}(z) \frac{z}{H(z)} \right]^{1/3} ,
\end{equation}
where the redshift at the drag epoch, $z_{\textrm{d}}$, is given by the fitting formula \cite{Eisenstein:1997ik}
\begin{equation}
z_{\textrm{d}}=\frac{1291(\Omega_{\textrm{m}} h^2)^{0.251}}{1+0.659(\Omega_{\textrm{m}} h^2)^{0.828}}[1+b_1(\Omega_{\textrm{b}} h^2)^{b_2}],
\end{equation}
where 
\begin{equation}
b_1=0.313(\Omega_{\textrm{m}} h^2)^{-0.419}[1+0.607(\Omega_{\textrm{m}} h^2)^{0.674}],
\end{equation}
and 
\begin{equation}
b_2=0.238(\Omega_{\textrm{m}} h^2)^{0.223}.
\end{equation}

To infer the cosmological parameters, BAO data can be used as constraints beside other surveys such as SNIa and CMB, in general the $\chi^{2}$ statistics is used for that purpose, and BAO contribution takes the from 
\begin{equation}
\bf \chi_{\textrm{BAO}}^{2}=X_{\textrm{BAO}}^{T}.C_{\textrm{BAO}}^{-1}.X_{\textrm{BAO}},
\end{equation}
where $\bf X_{\textrm{BAO}}$ is the difference vector between theoretical predictions (the third column of the tabe  \ref{tab_BAO} ) and observational measurements (the fourth column of the same table) and $C_{\textrm{BAO}}$ is the covariance matrix  given for the correlated data.

\begin{table}[H]
\begin{center}
\begin{tabular}{|c|c|c|c|c|c|c|}
\hline   
  \bf BAO name          &\bf$z$  &\bf BAO expression &\bf BAO measurement &$\bf {\sigma_{BAO}}$  &$\bf r_{s}^{\textit{fid}}$  &$\bf Ref$ \\
\hline
\hline
\multirow{1}{*} {6dFGS} &$0.106$ &$\frac{r_s}{D_V}$  &$0.327$   &$0.015$  &$-$   &\cite{Beutler:2011hx}       \\ [0.2cm]

 \hline                                       
\multirow{1}{*} {SDSS DR7 MGS} &$0.15$ &$D_V\frac{r_{s}^{fid}}{r_{s}}$ &$4.47$    &$0.16$    &$148.69$ &\cite{Ross:2014qpa} \\ [0.2cm] 

\hline 
                                             
\multirow{1}{*} {BOSS-LOWZ}    &$0.32$  &$D_V\frac{r_{s}^{fid}}{r_{s}}$    &$8.47$    &$0.17$ &$149.28$  &\cite{Anderson:2013zyy} \\ [0.2cm]                                         
\hline
\multirow{1}{*} {BOSS-CMASS}    &$0.57$  &$D_V\frac{r_{s}^{fid}}{r_{s}}$   &$13.77$    &$0.13$ &$149.28$  &\cite{Anderson:2013zyy} \\ [0.2cm]      
\hline               
 \multirow{3}{*} {WiggleZ}     &$0.44$  &   &$1716$    &$83$ & &    \\ [0.3cm] 
                               &$0.60$  &$D_V\frac{r_{s}^{fid}}{r_{s}}$    &$2221$ &$101$  &$148.6$    &\cite{Kazin:2014qga}\\[0.2cm]
                                       
                               &$0.73$  &  &$2516$ &$86$   &   &\\[0.3cm]                                        
                                        
\hline  

\multirow{3}{*} {}            &$0.38$  &$D_A(1+z)\frac{r_{s}^{fid}}{r_{s}}$  &$1512.39$   &$25.00$  & &             \\

\multirow{3}{*} {BOSS-DR12}   &     &$H\frac{r_{s}^{fid}}{r_{s}}$ &$81.2087$ &$2.3683$ & & \\ \cline{2-5}
  
                             &$0.51$  &$D_A(1+z)\frac{r_{s}^{fid}}{r_{s}}$     &$1975.22 $ &$30.10$          &$147.78$       &\cite{Alam:2016hwk} \\[0.3cm]
                             
\multirow{3}{*} {} & &$H\frac{r_{s}^{fid}}{r_{s}}$ &$90.9029$ &$2.3288$ & & \\\cline{2-5}                                                                         
                             &$0.61$  &$D_A(1+z)\frac{r_{s}^{fid}}{r_{s}}$      &$2306.68 $     &$37.08$      &            &\\[0.3cm]
\multirow{3}{*} {} & &$H\frac{r_{s}^{fid}}{r_{s}}$ &$98.9647$ &$2.5019$ & & \\                              
                                                                               
\hline
\hline
                                    
\end{tabular}

\caption{Summary of the Baryon Acoustic Oscillations data used in the current work.}\label{tab_BAO}

\end{center}
\end{table}
We should mention that WiggleZ and BOSS-DR12 data are correlated, and their covariance matrices are given respectively \cite{Kazin:2014qga,Alam:2016hwk} \\
\begin{equation}
\bf C_{\textrm{WiggleZ}}=10^{-4}\times\left( 
\begin{array}{cccc}
&2.17898878 & -1.11633321 & 0.46982851 \\ 
&           & 1.70712004            & -0.71847155\\ 
&           &                       & 1.65283175
\end{array} \right), 
\end{equation}
\begin{equation}
\bf C_{\textrm{BOSS-DR12}}=\left( 
\begin{array}{ccccccc}
&624.707 & 23.729 &325.332 &8.34963 &157.386 &3.57778 \\ 
&        &5.60873 &11.6429 &2.33996 &6.39263 &0.968056 \\ 
&        &        &905.777 &29.3392 &515.271 &14.1013 \\
&        &        &        &5.42327 &16.1422 &2.85334 \\
&        &        &        &        &1375.12 &40.4327 \\
&        &        &        &        &        &6.25936 \\
\end{array} \right). 
\end{equation}
The total $\bf \chi_{BAO}^2$ can be expressed as:
\begin{equation}
\bf \chi_{BAO}^2=\bf \chi_{6dFGS}^2+\chi_{SDSS}^2+\chi_{BOSS-LOWZ}^2+\chi_{BOSS-CMASS}^2 +\chi_{WiggleZ}^2 +\chi_{BOSS-DR12}^2 .
\end{equation}

\subsection{The $H(z)$ measurements}

In our analysis we have induced the Hubble expansion rate data to have a tighter constraints on our DE models, in general the $H(z)$ data can be derived either by the clustering of galaxies and quasars by measuring the BAO in the radial direction \cite{Gaztanaga:2008xz} or by the differential age method by expressing the Hubble parameter as
\begin{equation}
H\left(z\right)=-\frac{1}{\left(1+z\right)}\frac{dz}{dt}
\end{equation}
 and inferring $dz/dt$ from $\Delta z/\Delta t$ \cite{Jimenez:2001gg}, where $\Delta z$ and $\Delta t$ are respectively the redshift difference and the age difference between two passively evolving galaxies. In the current analysis we used a compilation of 36 data points of the Hubble parameter shown in table \ref{Hdata} where each data point is given with its corresponding reference.\\

While the Hubble parameter data points are not correlated, the $\chi_{H(z)}^2$ function can be written as

\begin{equation}
\chi^{2}_{H(z)}=\sum_{i=1}^{36} \left[ \frac{H_{obs,i}-H(z_{i})}{\sigma _{H,i}} \right]^{2} ,
\end{equation}
where $H_{obs,i}$ is the observational value of the Hubble parameter given for each redshift $z_{i}$ in the table \ref{Hdata} and $H(z)$ is the theoretical prediction of the Hubble parameter.
  \begin{table} [H] \label{Hdata}
\centering
\begin{tabularx}{220pt}{ cccccccc}
\toprule
\toprule
	{\centering{\bm{$z$}}}
	& \bm{$H(z)$}
    &\bm{$\sigma_H$}
	& {\bf Ref.}
	&\bm{$z$}
	& \bm{$H(z)$}
    &\bm{$\sigma_H$}
	& {\bf Ref.}
\\\midrule
	$0.07\phantom{0}$
	& $69.0 \phantom{}$
	&$19.6$
	& \cite{Zhang:2012mp} 
	&$0.48$
	&$97.0$
	&$62.0$
	&\cite{Stern:2009ep} 
 \\\midrule
	$0.09\phantom{0}$
	& $69.0 \phantom{}$
	&$12.0$
	& \cite{Stern:2009ep} 
	&$0.57$
	&$96.8$
	&$3.4$
	&\cite{Anderson:2013zyy}
 \\\midrule
	$0.12\phantom{0}$
	& $68.6\phantom{}$
	&$26.2$
	& \cite{Zhang:2012mp} 
	&$0.593$
	&$104.0$
	&$13.0$
	&\cite{Moresco:2012jh}
 \\\midrule
	$0.17\phantom{0}$
	& $83.0 \phantom{}$
	&$8.0$
	& \cite{Stern:2009ep} 
	&$0.60$
	&$87.9$
	&$6.1$
	& \cite{Stern:2009ep} 
 \\\midrule
	$0.179\phantom{0}$
	& $ 75.0\phantom{0}$
	&$4.0$
	&\cite{Moresco:2012jh} 
	&$0.68$
	&$92.0$
	&$8.0$
	&\cite{Moresco:2012jh}
 \\\midrule
	$0.199\phantom{0}$
	& $ 75.0 \phantom{0}$
	&$ 5.0$
	& \cite{Moresco:2012jh} 
	&$0.73$
	&$97.3$
	&$7.0$
	& \cite{Stern:2009ep} 
 \\\midrule
	$0.2\phantom{0}$
	& $72.9\phantom{}$
	&$ 29.6$
	& \cite{Zhang:2012mp} 
	&$0.781$
	&$105.0$
	&$12.0$
	&\cite{Moresco:2012jh}
 \\\midrule
	$0.27\phantom{0}$
	& $77.01$
	&$14.0$
	& \cite{Stern:2009ep} 
	&$0.875$
	&$125.0$
	&$17.0$
	&\cite{Moresco:2012jh}
 \\\midrule
	$0.28\phantom{0}$
	& $88.8$
	&$36.6$
	& \cite{Zhang:2012mp} 
	&$0.88$
	&$90.0$
	&$40.0$
	& \cite{Stern:2009ep} 
 \\\midrule
	$0.35\phantom{0}$
	& $82.7$
	&$8.4$
	& \cite{Chuang:2012qt} 
	&$0.9$
	&$117.0$
	&$23.0$
	& \cite{Stern:2009ep} 
 \\\midrule
	$0.352\phantom{0}$
	& $83.0$
	&$14.0$
	& \cite{Moresco:2012jh} 
	&$1.037$
	&$154.0$
	&$20.0$
	&\cite{Moresco:2012jh}
 \\\midrule
	$0.3802\phantom{0}$
	& $83.0$
	&$13.5$
	& \cite{Moresco:2016mzx}
	&$1.3$
	&$168.0$
	&$17.0$
	& \cite{Stern:2009ep} 
 \\\midrule
	$0.4\phantom{0}$
	& $95.0$
	&$17.0$
	& \cite{Stern:2009ep} 
	&$1.363$
	&$160.0$
	&$33.6$
	&\cite{Moresco:2015cya}
 \\\midrule
	$0.4004\phantom{0}$
	& $77.0$
	&$10.2$
	& \cite{Moresco:2016mzx}
	&$1.43$
	&$177.0$
	&$18.0$
	& \cite{Stern:2009ep} 
 \\\midrule
	$0.4247\phantom{0}$
	& $87.1$
	&$11.2$
	& \cite{Moresco:2016mzx}
	&$1.53$
	&$140.0$
	&$14.0$
	& \cite{Stern:2009ep} 
 \\\midrule
	$0.44\phantom{0}$
	& $82.6$
	&$7.8$
	& \cite{Blake:2012pj} 
	&$1.75$
	&$202.0$
	&$40.0$
	& \cite{Stern:2009ep} 
 \\\midrule
	$0.44497\phantom{0}$
	& $92.8$
	&$12.9$
	& \cite{Moresco:2016mzx}
	&$1.965$
	&$186.5$
	&$50.4$
	&\cite{Moresco:2015cya}
 \\\midrule
	$0.4783\phantom{0}$
	& $80.9$
	&$9.0$
	& \cite{Moresco:2016mzx}
	&$2.34$
	&$222.0$
	&$7.0$
	&\cite{Stocker:2018avm}
 \\
\bottomrule
\bottomrule
\end{tabularx}
\caption{This table shows the measurements of the Hubble expansion.$ H(z)$ data used in the current analysis are in the unit of  km $\textrm{s}^{-1}$ $\textrm{Mpc}^{-1}$.}\label{Hdata}
\end{table}

Finally, the $\chi_{tot}^2$ is the sum of all the $\chi^2$ previously defined:
\begin{equation}
\bf \chi_{tot}^2=\bf \chi_{SN}^2+\chi_{\textrm{CMB}}^2+\chi_{BAO}^2+\chi_{H(z)}^2.
\end{equation}

\section{Dark energy models}\label{DEmodels}
\subsection{WCDM model}

A BR singularity labeled in this work as the model A can be induced  by a perfect fluid whose EoS parameter is  constant and smaller than -1  
\begin{equation}
p_{\textrm{d}}=w_{\textrm{d}}\rho_{\textrm{d}},
\end{equation}
where $w_{\textrm{d}}$ is the EoS parameter. Therefore, given the previous EoS and using the conservation equation, the energy density evolves with the scale factor as 
\begin{equation}
\rho_{\textrm{d}}(a)=\rho_{\textrm{d}0}a^{-3(1+w_{\textrm{d}})}, 
\end{equation}
where $\rho_{\textrm{d}0}$ is the current value of DE density while at present the scale factor is set to be $a_0=1$. Finally, the Friedmann equation can be expressed as follows
\begin{equation}
E(a)^2=\Omega_{\textrm{r}} a^{-4}+\Omega_{\textrm{m}} a^{-3}+\Omega_{\textrm{d}} a^{-3(1+w_{\textrm{d}})}.
\end{equation}

We have redefined the Hubble parameter as $E\left(a\right)\equiv H\left(a\right)/H_{0}$, where $H_0$ is the value of the Hubble parameter at present. In the same way, $\Omega_{\textrm{r}}$, $\Omega_{\textrm{m}}$ and $\Omega_{\textrm{d}}$ represent the current fractional energy densities of radiation, matter and dark energy, respectively. Therefore, once $\Omega_{\textrm{r}}$ and $\Omega_{\textrm{m}}$  are fixed, the value of $\Omega_{\textrm{d}}$ should be settled as $\Omega_{\textrm{d}}=1-\Omega_{\textrm{m}}-\Omega_{\textrm{r}}$.  As the fit also includes  the value of  $\Omega_{\textrm{m}}$, which is model dependent,  we have imposed the  value of $8\times 10^{-5}$ to $\Omega_{\textrm{r}}$ for all models.

\subsection{ LR abrupt event model}

This model labeled as the model B, is characterised by inducing a future abrupt event known as a LR. The EoS that relate the pressure and the energy density can be written as  \cite{Nojiri:2005sx,Stefancic:2004kb}
\begin{equation}
p_{\textrm{d}}=-\left( \rho_{\textrm{d}}+\beta\rho_{\textrm{d}}^{\frac{1}{2}}\right), 
\end{equation}
where $B$ is a positive  constant. Using the conservation equation together with the  EoS,  the energy density in terms of the scale factor is given by 
\begin{equation}
\rho_{\textrm{d}}(a)=\rho_{\textrm{d}0}\left[ \frac{3}{2}\frac{\beta}{\sqrt{\rho_{\textrm{d}0}}}\ln\left(a\right)+1 \right]^2 \hspace{0.2cm} ,
\end{equation}
Then, solving the Friedmann equation, the evolution of the Hubble parameter can be expressed as follows
\begin{equation}\label{a}
E^{2}(a)=\Omega_{\textrm{r}} a^{-4}+\Omega_{\textrm{m}} a^{-3}+\Omega_{\textrm{d}}\left(1+\frac{3}{2}\sqrt{\frac{\Omega_{\textrm{lr}}}{\Omega_{\textrm{d}}}} \ln(a)\right)^2 , 
\end{equation}
In order to get  a compact expression for $E^{2}(a)$ we have redefined the dimensionless quantity $\Omega_{\textrm{lr}}$ as $\Omega_{\textrm{lr}}\equiv 8\pi G/3H_{0}^{2}\beta^{2}$.

\subsection{LSBR abrupt event model}

The model C that induces the future abrupt event known as LSBR have an  EoS as follows \cite{Bouhmadi-Lopez:2014cca}
\begin{equation}
p_{\textrm{d}}=-\left( \rho_{\textrm{d}}+\frac{\alpha}{3}\right) ,
\end{equation}
As can be seen, this model differs from the widely known $\Lambda$CDM model by the parameter $A$ which is a positive constant.  Therefore, using the conservation equation together with the EoS, the energy density evolves with the scale factor as 
\begin{equation}
\rho_{\textrm{d}}(a)=\rho_{\textrm{d}0}+\alpha\ln\left( a\right).
\end{equation}
Therefore, after solving the Friedmann equation, we get the evolution of the Hubble parameter in terms of the scale factor
\begin{equation}
E^{2}(a)=\Omega_{\textrm{r}} a^{-4}+\Omega_{\textrm{m}} a^{-3}\\
+\Omega_{\textrm{d}}\left( 1-\frac{\Omega_{\textrm{lsbr}}}{\Omega_{\textrm{d}}}\ln(a)\right),
\end{equation}
where we have redefined the dimensionless quantity $\Omega_{\textrm{lsbr}}$ as $\Omega_{\textrm{lsbr}}\equiv 8\pi G/3H_{0}^{2}\alpha$ in order to compactify the notation and present the results in a simpler way.

\section{background results}\label{results}

\begin{figure}[H]
\includegraphics[scale=1.1]{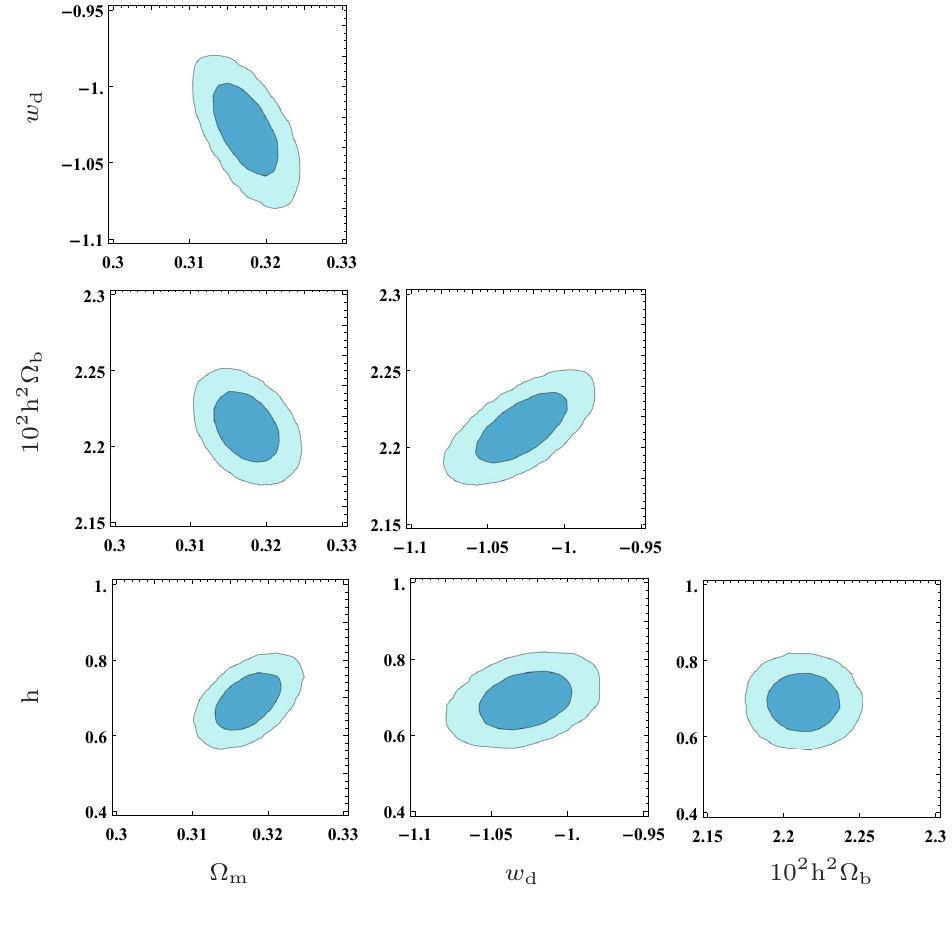}
\caption{ These figures correspond to 1$\sigma$ and 2$\sigma$ confidence contours obtained from SNIa+CMB+BAO+$\textrm{H(z)}$ data for the model A. }\label{contourBR}
\end{figure}

\begin{figure}[H]
\includegraphics[scale=1.1]{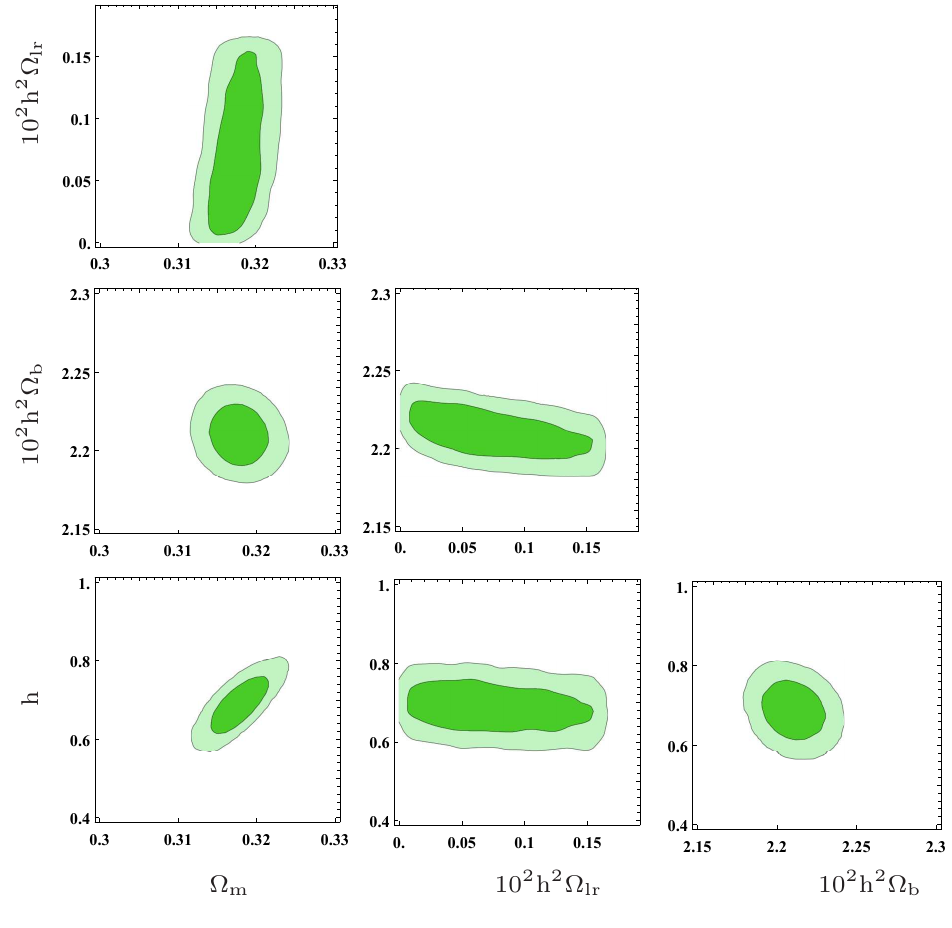} 

\caption{  These figures correspond to 1$\sigma$ and 2$\sigma$ confidence contours obtained from SNIa+CMB+BAO+$\textrm{H(z)}$ data for the model B. }\label{contourLR}
\end{figure}

\begin{figure}[H]
\includegraphics[scale=1.1]{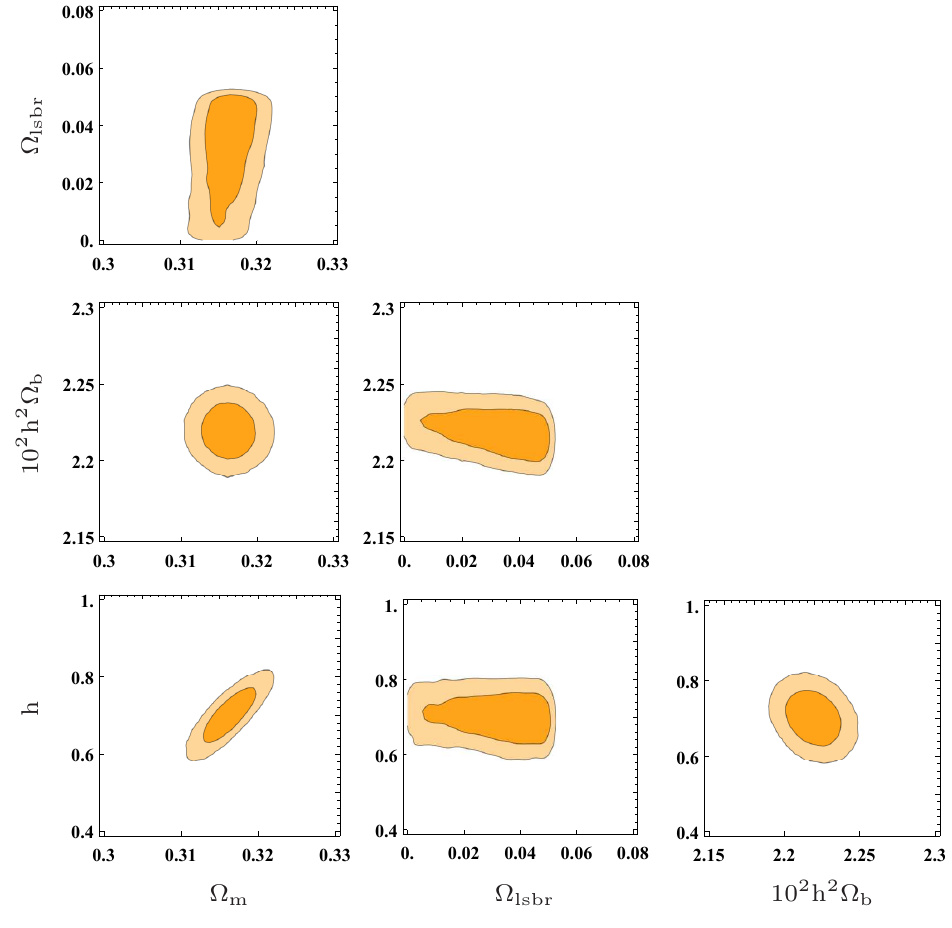} 
\caption{ These figures correspond to 1$\sigma$ and 2$\sigma$ confidence contours obtained from SNIa+CMB+BAO+$\textrm{H(z)}$ data for the model C.}\label{contourLSBR}
\end{figure}

In this section, we present the obtained results for the observational fit. The figure~\ref{contourBR}, \ref{contourLR} and \ref{contourLSBR} show the contour plots of the model parameters corresponding to the model A, B and C, respectively. The numerical results are all gathered in Table~\ref{tab1}

 The criteria methods used in the literature to compare between models are mainly the $\chi_{min}^{red}$ and the corrected Akaike Information Criterion ($AIC_c$) defined as \cite{AIC,Basilakos:2017rgc,Sagredo:2018rvc}
\begin{equation}\label{AIC}
AIC_c = -2\ln{\cal{L}}_{max}+2N_p+\frac{2N_p(N_p+1)}{N_d-N_p-1},
\end{equation}
where $N_p$ denotes the number of parameters and $N_d$ denotes the number of data.
In the case of Gaussian errors, $\chi^2_{min}=-2\ln{\cal{L}}_{max}$ and
\begin{equation}\label{AIC2}
AIC_c = \chi^2_{min}+2N_p+\frac{2N_p(N_p+1)}{N_d-N_p-1}.
\end{equation}

In practice,  we do not care about $AIC_c$ value itself in model comparisons. Actually we are interested to calculate the $\Delta AIC_c$ between models, i.e, $\Delta AIC_c =  AIC_{c,model}- AIC_{c,min}$. The model with a minimal value of $AIC_c$ is more favoured by data and it is chosen as a reference model. Roughly speaking, the models with $0 < \Delta \textrm{AIC}_c< 2$ have substantial support, those with $4 < \Delta \textrm{AIC}_c< 7$ have considerably less support, and models with $\Delta \textrm{AIC}_c> 10$ have essentially no support, with respect to the reference model. Finally, $\Lambda$CDM model is also favoured by the $\chi_{min}^{red}$ selection.
In Table~\ref{tab1}, we show the best fit and the mean values of the parameters. The free parameter vectors when assuming a flat Universe for $\Lambda$CDM, $A$, $B$, and $C$ models are respectively $\theta_{\Lambda CDM}=(\Omega_m,h,\Omega_bh^2)$, $\theta_{A}=(\Omega_m,\omega_d,h,\Omega_bh^2)$, $\theta_{B}=(\Omega_m,\Omega_{lr},h,\Omega_bh^2)$ and $\theta_{C}=(\Omega_m,\Omega_{lsbr},h,\Omega_bh^2)$. The $\chi_{tot}^{2}$ and ${\chi_{tot}^2}^{red}$ are also given in the same table. In order to study the statistical significance of our constraints, we compute  $AIC_c$ and $\Delta AIC_c$ with $N_d=1097$, $N_p=3$ for $\Lambda$CDM and $N_p=4$ for the rest of the models. The values of $\Delta AIC_{c}$ are $2.104677$, $2.124677$ and $2.154677$ for the models A, B and C respectively, and are given for the purpose of models comparison. Given that all the $\Delta AIC_c$ values are close to $\sim 2$, it makes the three models A, B and C competitive and supported by the data. But still the strongly favoured model is the $\Lambda$CDM.

\begin{table}[H]
\begin{center}
\begin{tabular}{cccccccc}
\hline   
                   \bf Model              &\bf Par                &\bf Best fit     &\bf Mean               &$\bf {{\chi}^2_{tot}}$  &$\bf {{\chi}^2_{tot}}^{red}$ &$AIC_c$  &$\Delta AIC_c$ \\
\hline
\hline
\multirow{3}{*} {$\Lambda$CDM}                  &$\Omega_{\textrm{m}}$              &$0.318349^{+0.00248001}_{-0.00248001}$  &$0.31834^{+0.00248987}_{-0.00248987}$   &$1047.42$  &$0.957422$   &$1053.441953$ &0         \\  
                                                                 
                                       &$h$                     &$0.69814_{-0.0480814}^{+0.0480814} $      &$0.698602_{-0.0481787}^{+0.0481787} $     & \\[0.1cm]
                                        &$\Omega_{\textrm{b}} h^{2}$        &$0.022218_{-0.000120872}^{+0.000120872}$    &$0.0222202_{-0.000122619}^{+0.000122619}$       &\\[0.1cm]                                                         
 \hline                                       
\multirow{3}{*} {BR}                    &$\Omega_{\textrm{m}}$              &$0.317173_{-0.00318473}^{+0.00318473}$    &$0.317327_{-0.0031808}^{+0.0031808}$    &$1047.51$ &$0.958380$  &$1055.54663$ &$2.104677$                 \\ [0.1cm] 
                                        &$w_{\textrm{br}}$          &$-1.02758_{-0.0240102}^{+0.0240102}$     &$-1.02874_{-0.0239306}^{+0.0239306}$     &\\[0.1cm]
                                       
                                        &$h$                     &$0.691013_{-0.0507771}^{+0.0507771} $     &$0.691523_{-0.0507536}^{+0.0507536} $       &\\[0.1cm]
                                        &$\Omega_{\textrm{b}} h^{2}$        &$0.0221218_{-0.000170789}^{+0.000170789} $  &$0.022123_{-0.000170538}^{+0.000170538} $          &\\[0.1cm]                                        
                                        
\hline               
 \multirow{3}{*} {LR}                   &$\Omega_{\textrm{m}}$              &$0.317198_{-0.00276851}^{+0.00276851}$     &$0.317705_{-0.00280131}^{+0.00280131}$    &$1047.53$ &$0.958398$ &$1055.56663$ &$2.124677$   \\ [0.1cm] 
                                        &$\Omega_{\textrm{lr}}$           &$0.000445721_{-0.000416159}^{+0.000416159}$    &$0.000763824_{-0.000416359}^{+0.000416359}$ &\\[0.1cm]
                                       
                                        &$h$                     &$0.694604_{-0.0494111}^{+0.0494111} $   &$0.688584_{-0.0493315}^{+0.0493315} $        &\\[0.1cm]
                                        &$\Omega_{\textrm{b}} h^{2}$        &$0.0221295_{-0.000130585}^{+0.000130585} $  &$0.0221028_{-0.000132755}^{+0.000132755} $          &\\[0.1cm]                                        
                                        
\hline  

\multirow{3}{*} {LSBR}                  &$\Omega_{\textrm{m}}$              &$0.317115^{+0.00253975}_{-0.00253975}$  &$0.316144^{+0.00253899}_{-0.00253899}$   &$1047.56$  &$0.958426$ &$1055.59663$ &$2.154677$             \\  
                                        &$\Omega_{\textrm{lsbr}}$         &$0.0500261_{-0.0130141}^{+0.0130141} $     &$0.0299424_{-0.0133398}^{+0.0133398} $ &          &\\[0.1cm]
                                       
                                        &$h$                     &$0.695705_{-0.0481201}^{+0.0481201} $      &$0.701962_{-0.0481465}^{+0.0481465} $     & \\[0.1cm]
                                        &$\Omega_{\textrm{b}} h^{2}$        &$0.022138_{-0.000121724}^{+0.000121724}$    &$0.0221928_{-0.000121811}^{+0.000121811}$       &\\[0.1cm]                                        
\hline
\hline
                                    
\end{tabular}

\caption{Summary of the best fit and the mean values of the cosmological parameters.}\label{tab1}

\end{center}
\end{table}


\section{Perturbation Results}\label{pertresults}
Before tackling cosmological perturbations of the models introduced in Sec.~\ref{DEmodels}, we show how the EoS parameter evolves in time for the models A, B and C defined in the above mentioned Sec.~\ref{DEmodels}. The reason for highlighting $w_{\textrm{d}}$ for these models, is the important role they play for the initial condition of DE perturbations (cf. Eqs. (3.20) and (3.21) of Ref. \cite{Albarran:2016mdu})

\begin{figure}[h]
 \includegraphics[width=12cm]{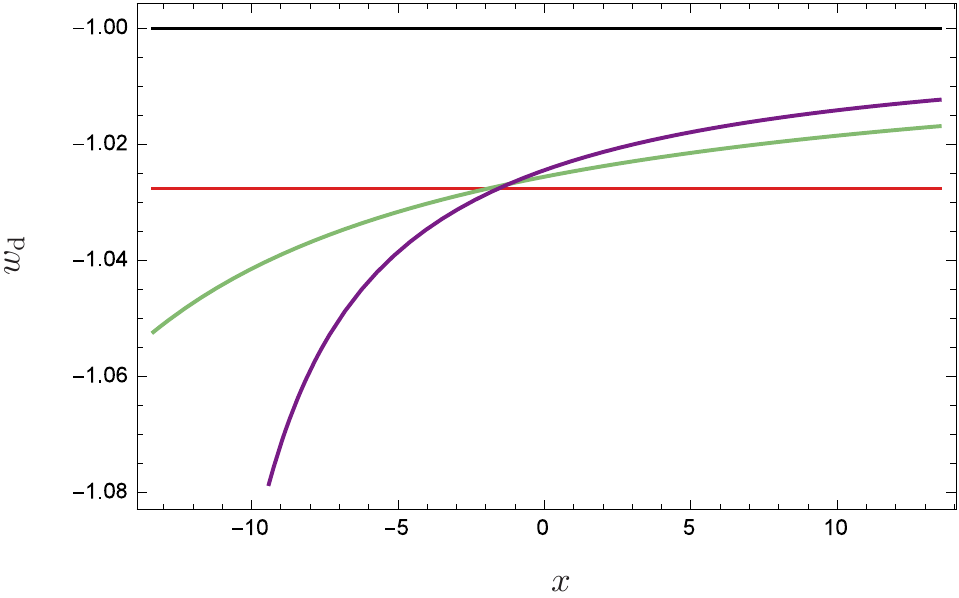}
\caption{This figure shows the evolution of the EoS parameter of DE, $w_\textrm{d}$, versus $x\equiv\ln\left(a\right)$ for the model parameters given by the best fit in table~\ref{tab1}. The model A corresponds with the red line describing a constant EoS. The model B is represented by the green line while the model C is shown in purple. The $\Lambda$CDM model is presented as a  black line at $w_\textrm{d}=-1$.}\label{wevol}
\end{figure}

In order to get Fig. \ref{wevol}, we use the best fit parameters values obtained in Sec.~\ref{results} which are shown in table~\ref{tab1}. In this figure, the black line corresponds to $\Lambda$CDM, the red line  to model A; i.e. a constant EoS and smaller than $-1$, and leading to a BR, the green line to model B; i.e. the one leading to a LR, and the purple line to model C; i.e. the one leading to a LSBR.

\begin{figure}[H]
 \includegraphics[width=\textwidth]{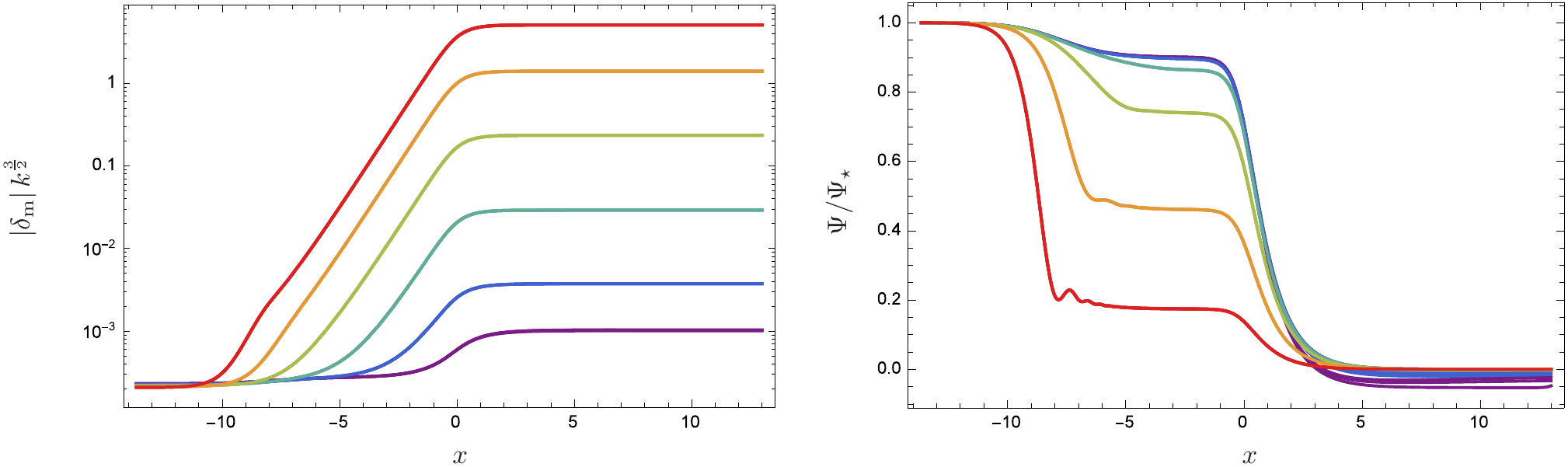}
\caption{The left panel of this figure shows the evolution of the mater density contrast while the right panel shows the evolution of the gravitational potential divided by its initial value. Both panels show a  perfect overlapping of the results corresponding to  the models A, B, C and $\Lambda$CDM. The results corresponding to a given mode are represented by a particular color as follows: $k_1$ (purple), $k_2$ (dark blue), $k_3$ (light blue), $k_4$ (green), $k_5$ (orange) and  $k_6$ (red).}\label{best_deltam_Psi}
\end{figure}

We next show our results for the evolution of the cosmological perturbations of radiation, DM and DE. We have computed these perturbations for six relevant modes which run from roughly a mode corresponding  to the current Hubble horizon  ${k_1=3.33\times10^{-4}\textrm{h Mpc}^{-1}}$ to the largest mode where the linear regime is still valid, $k_6=0.1 \ \textrm{h Mpc}^{-1}$. The six modes are equidistant in a logarithmic scale as follows
\begin{equation}
k_{j}=k_1\left(\frac{k_{6}}{k_{1}}\right)^{\frac{j-1}{5}}, 
\end{equation}\label{kmodes}
where $i$ runs from $1$ to $6$. Therefore, the numerical value of the six modes are
\begin{align}
&{k_{1}=3.33\times10^{-4}\textrm{h Mpc}^{-1}}, \qquad &k_{4}=1.02\times10^{-2}\textrm{h Mpc}^{-1},\\
&k_{2}=1.04\times10^{-4}\textrm{h Mpc}^{-1}, \qquad &k_{5}=3.19\times10^{-2}\textrm{h Mpc}^{-1},\\
&k_{3}=3.26\times10^{-3}\textrm{h Mpc}^{-1}, \qquad &k_{6}=1.00\times10^{-1}\textrm{h Mpc}^{-1}.
\end{align}
As mentioned on the previous section, the evolution of  the perturbations  are computed from well inside the radiation dominated epoch, $x_\textrm{i}=-13.8$, till the distant future, $x_\textrm{f}=13$, where the DE completely dominates the dynamics of the Universe. We consider $x_\textrm{f}$ large enough to detect relative deviations between the studied models. We next present the main results.
The left panel of figure~\ref{best_deltam_Psi} shows the evolution of the matter density contrast of models A, B and C together with $\Lambda$CDM. As can be seen, there is no significant deviation since all the modes show almost identical evolution. As expected, the matter density contrast of each mode grows linearly when the mode enters the horizon and reaches it maximum value when DE starts dominating. 

\begin{figure}[H]
 \includegraphics[width=\textwidth]{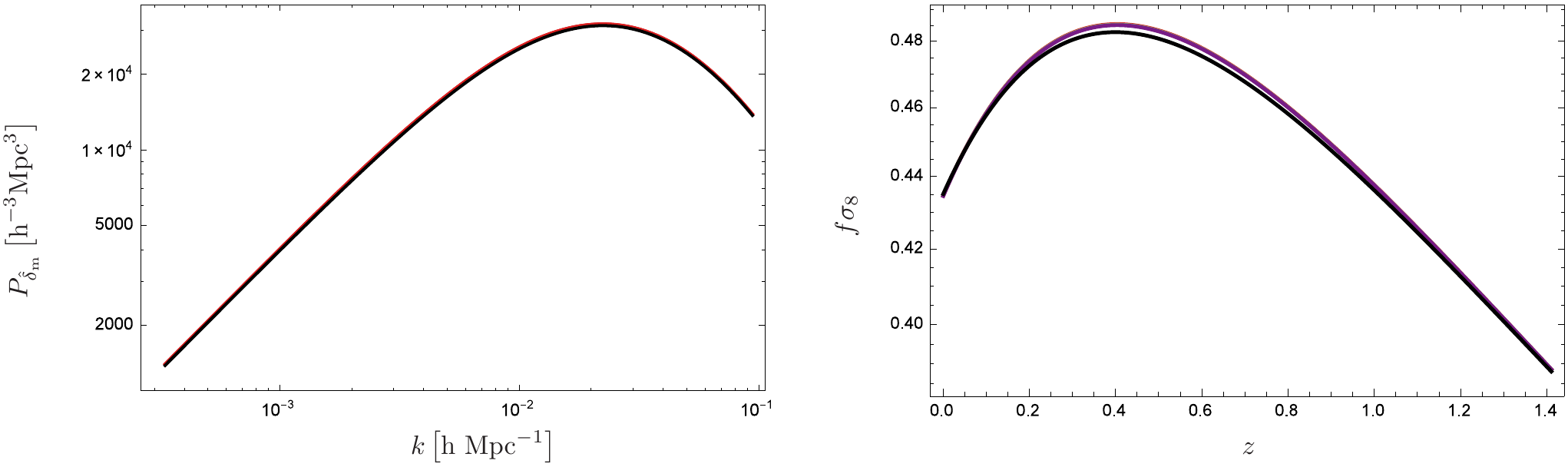}
\caption{The left panel of this figure shows the matter power spectrum when using the model parameters given table \ref{tab1}. The result corresponding to the models A, B and C are shown overlapped in red, while the black color corresponds to the $\Lambda$CDM model. As can be seen, both results are almost indistinguishable. On the other hand, the right panel of this figure shows the evolution of the $f\sigma_{8}$. The evolution corresponding to the $\Lambda$CDM model is shown in black color, while the results given by  the models A, B and C are gathered in a single purple curve.}\label{best_mps_fs8}
\end{figure}

The right panel of figure~\ref{best_deltam_Psi} shows the evolution of the gravitational potential, $\Psi$,  divided by its initial value, $\Psi_{*}$. The results of the models A, B, C and $\Lambda$CDM are plotted together in the figure. Once again, the overlap is almost perfect, except for the small deviations presented by all the modes at very large scales, we will discuss
 this feature later on.
 The left panel of figure~\ref{best_mps_fs8} shows the predicted  current matter power spectrum of  models A, B, C and $\Lambda$CDM. The black curve corresponds to the  $\Lambda$CDM model while the  models A, B and C are shown overlapped in a single red curve. The overlap is almost perfect being impossible to distinguish any footprints between these models and $\Lambda$CDM. In general, the main behaviour is in accordance with that found in the literature and gives a suitable description of the current matter power spectrum.

The right panel of figure~\ref{best_mps_fs8} shows the evolution of $f\sigma_{8}$. The results of the models A, B and C are overlapped and appear as purple curve, while the results corresponding to $\Lambda$CDM are in black. There are no significant deviations between  models A, B and C. However, there is some deviation with regards to $\Lambda$CDM for $z\sim0.3$ to $z\sim0.6$. This result implies that $f\sigma_{8}$ is larger for phantom DE models as compared with  $\Lambda$CDM. This result is in agreement with \cite{Albarran:2016mdu,Albarran:2017kzf}.

As can be seen from figures~\ref{best_deltam_Psi} and \ref{best_mps_fs8}, it is very difficult to distinguish the models A, B and C as no significant deviation is observed on the matter density contrast and gravitational potential. In view of this, we find convenient to compute relative deviations with respect to $\Lambda$CDM.

\begin{figure}[H]
 \includegraphics[width=\textwidth]{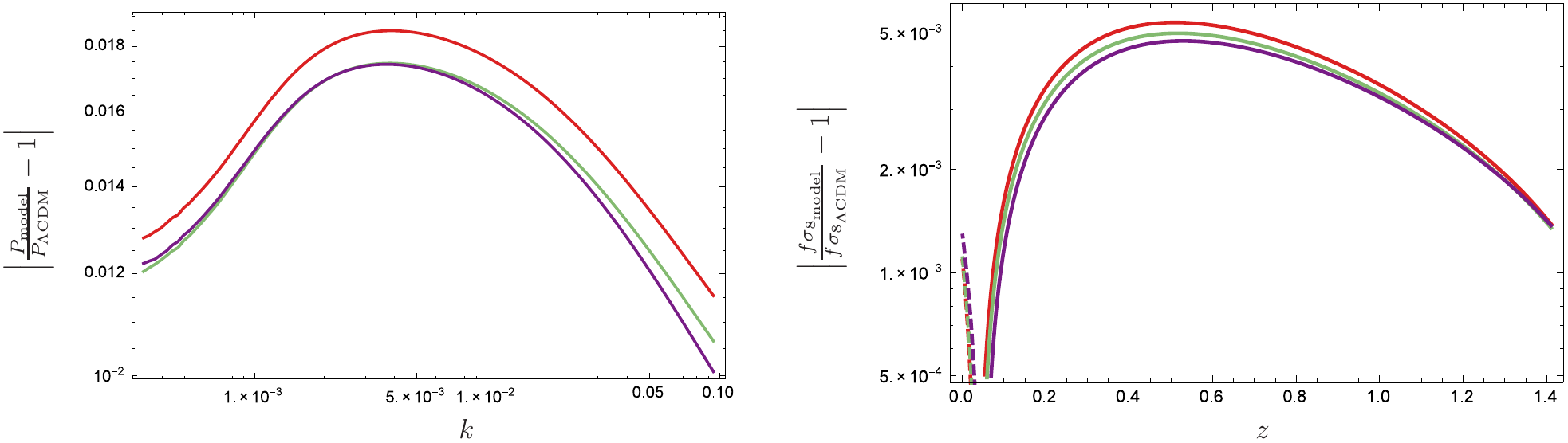}
\caption{The left and right panels of this figure show the relative deviation  of the matter power spectrum  and $f\sigma_{8}$ with respect to $\Lambda$CDM, respectively. The colors red, green and purple have been used to plot results corresponding to the models A, B and C, respectively. Both plots are drawn in a  logarithmic scale, where the dashed line is used to denote negative values and the solid line to denote positive values. }\label{best_rel_mps_fs8}
\end{figure}

 The left panel of figure~\ref{best_rel_mps_fs8} shows the relative difference  for the  models A, B and C with respect to $\Lambda$CDM of the matter power spectrum. As can be seen,  all the models show a very similar behaviour. The deviation is positive for all the modes, reaching a maximum at $k\simeq 5\times10^{-3}\textrm{h Mpc}^{-1}$  where the largest deviation is around a 2\%.   The right panel of Figure~\ref{best_rel_mps_fs8} shows the relative difference of $f\sigma_{8}$ for the  models A, B and C with respect to $\Lambda$CDM.  Such deviations  show a maximum at $z\simeq0.5$ with a value around 0.5\%.  The deviation of $f\sigma_{8}$ with respect to $\Lambda$CDM is positive for all the redshift range except for the lowest values. In fact, a  transition to a negative difference is observed around redshift $z\simeq0.05$ . This later deviation increases as we approach the present. Once again, we obtain very similar plots when comparing the results obtained for the models A, B and C. We find that the largest deviations correspond to the model A and the smallest one to the model C.

\begin{figure}[H]
 \includegraphics[width=\textwidth]{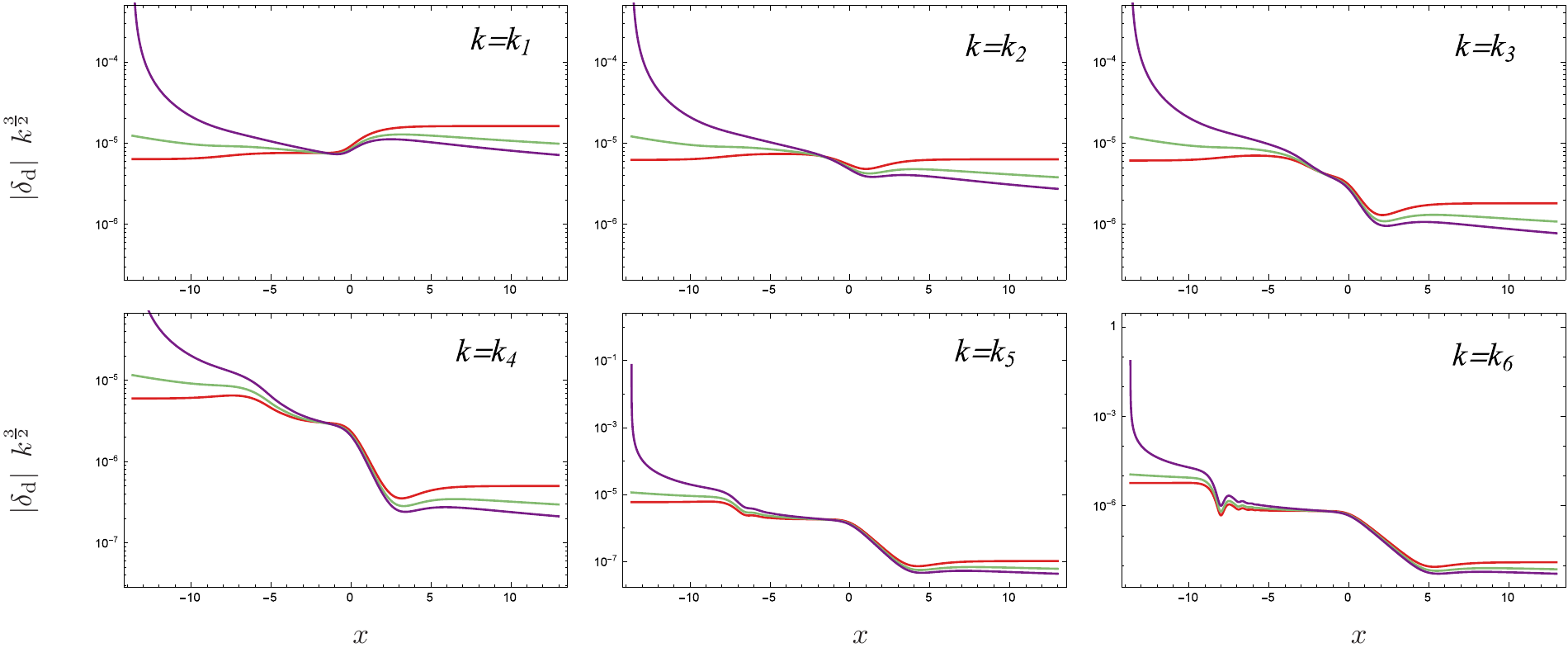}
\caption{This figure shows the DE density contrast for the models A (red), B (green) and C (purple). Each panel corresponds to a specific wave-number $k$.}\label{best_deltad}
\end{figure}

Figure~\ref{best_deltad} shows  the evolution of  DE  density contrast, where each panel corresponds to a given mode. As the perturbations of DE into the $\Lambda$CDM model vanishes, we do not compare them with the result of our models. The initial values of the perturbations  $\delta_{\textrm{d}}$ are not significantly affected for different modes, i.e. $k$. However, they depend on the specific model because the EoS of DE is model dependent, in particular at $x=x_{\textrm{i}}$ where we start our numerical integrations. Moreover, given that the EoS parameter for the model C is closer to $-1$ and it shows the highest density contrast for DE at earlier time (cf. Eq. (\ref{initcond1})). The model A presents the opposite behaviour, while the model B shows an intermediate behaviour. This hierarchical behaviour is inverted in the future, where $\delta_{\textrm{d}}$ gets larger values for the model A and smaller values for the model C. This transition occurs at very low redshifts which brings difficulties in distinguishing the different DE models analysed.

\begin{figure}[h]
 \includegraphics[width=\textwidth]{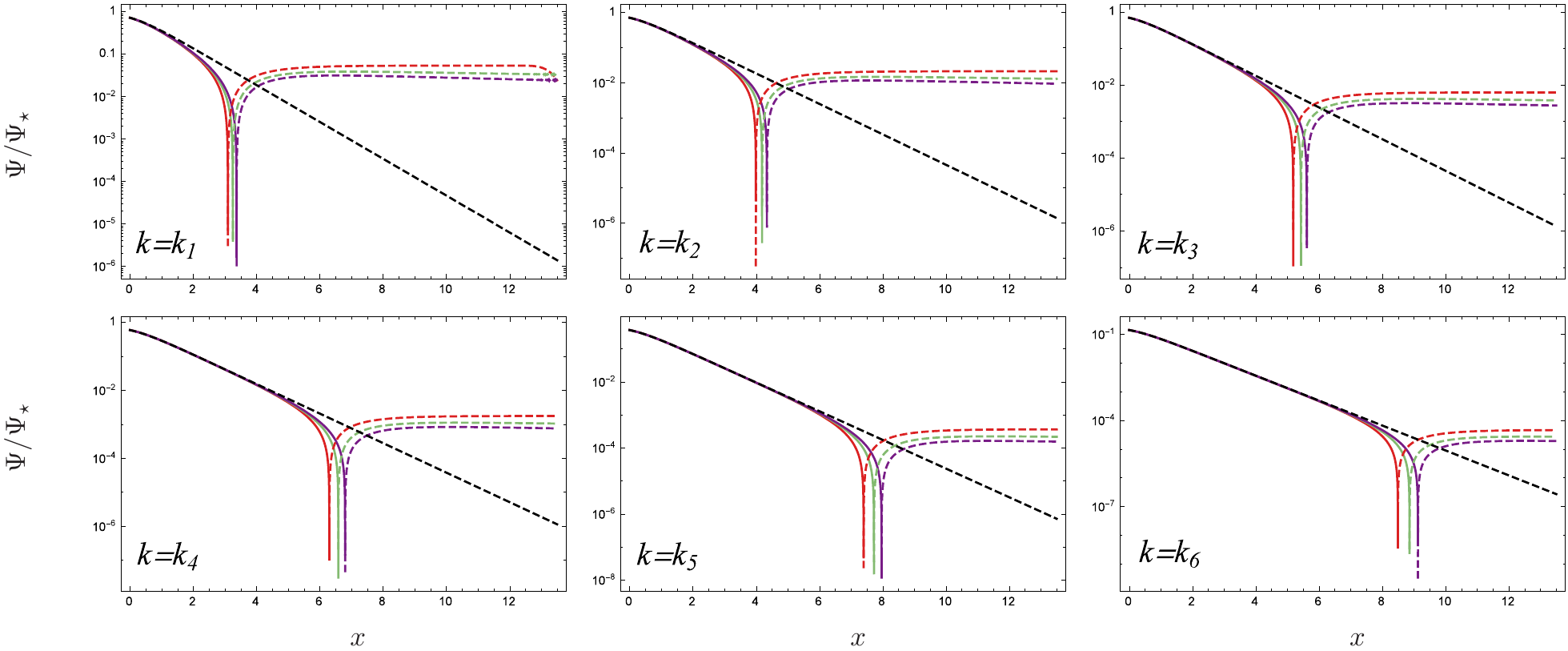}
\caption{This figure shows  the evolution of the gravitational potential, $\Psi$, divided by its initial value, $\Psi_{\star}$. As done in the previous figures, each panel corresponds to a given  value of wave-number while each color represents a particular model (model A red, model B  green and model C  purple). The dashed lines denote negative values while solid lines represent positive values.  The black dashed lines correspond to the $\Lambda$CDM model.}\label{best_Psi_final}
\end{figure}

We have seen that the difference between models are very small. However, we know very well that  each model induces a different and unique abrupt event in the far future. Therefore, we have focused on the evolution of the gravitational potential at very large scale factors.  Figure~\ref{best_Psi_final} shows  the evolution of the gravitational potential, $\Psi$, from the present cosmic time to the far future for different modes. As can be seen, at present the gravitational potential of  all the models are very close to  $\Lambda$CDM. However, at some point in the future, the gravitational potential flips its sign and evolves towards a constant negative value. Within these asymptotic evolutions, model  A introduces the highest deviation while model C introduces the smallest one. The flip on the sign of  $\Psi$ occurs sooner in the model A, then in the model B and finally in the model C, independently of the mode. On the other hand, the smallest is the mode the sooner occurs the sign flip on the gravitational potential. 

Finally, and based on the best parameters of table~\ref{tab1}, we have computed the reduced $\chi^{2}$ for the $f\sigma_{8}$ analysis  in order to have a numerical value that could quantify the difference between  models. Note that we do not run a different chain taking into account such $f\sigma_{8}$ data. We rather perform  a simple analysis in order to get some preliminary results involving background and perturbative observations. Here, we have used the ``Gold 2017'' data \cite{Nesseris:2017vor}. This data provides a set of the latest measurements of $f\sigma_{8}$ values\footnote{An extension of the  Gold 2017 compilation is given by \cite{Kazantzidis:2018rnb}.} (ranging from a redshift $z\sim0$ to $z\sim1.4$) where all  samples are considered as independent. The obtained  results are shown in table~\ref{chisqtable2}. As can be seen,  the model A is the model that more deviates from $\Lambda$CDM, followed by the model B and finally by the model C. The best value is still given by the $\Lambda$CDM model, however, among the studied phantom models in this work, the model C is observationally preferred.  We notice that the background classification of the models A, B and C, table \ref{tab1},  is not in agreement with that based on the measurement of  $f\sigma_{8}$. In a future work, we will carry a full Monte Carlo Markov Chain which takes into account the background and the perturbative quantities.

  \begin{table}[h]
\begin{tabular}{ |m{1cm}|m{2cm}| m{2cm}| }
\hline
 \centering Model &  \centering Event &  \centering $\chi^{2 }$ \tabularnewline
\hline
\centering {$\Lambda$CDM} & \centering {De Sitter} & \centering {1.13498}  \tabularnewline
\hline
 \centering  A &\centering  BR & \centering 1.16163  \tabularnewline
 \hline
  \centering  B & \centering LR & \centering 1.15919 \tabularnewline
 \hline
\centering  C & \centering LSBR & \centering 1.15680 \tabularnewline
 \hline
\end{tabular}
   \caption{This table presents the values of the reduced $\chi^{2}$ for each model. These results have been obtained using a data collection of independent survey known as ``Gold 2017'' growth dataset, which values are shown in table~\ref{dataset1}.}\label{chisqtable2}

\end{table}

\begin{figure}[h]
 \includegraphics[width=11cm]{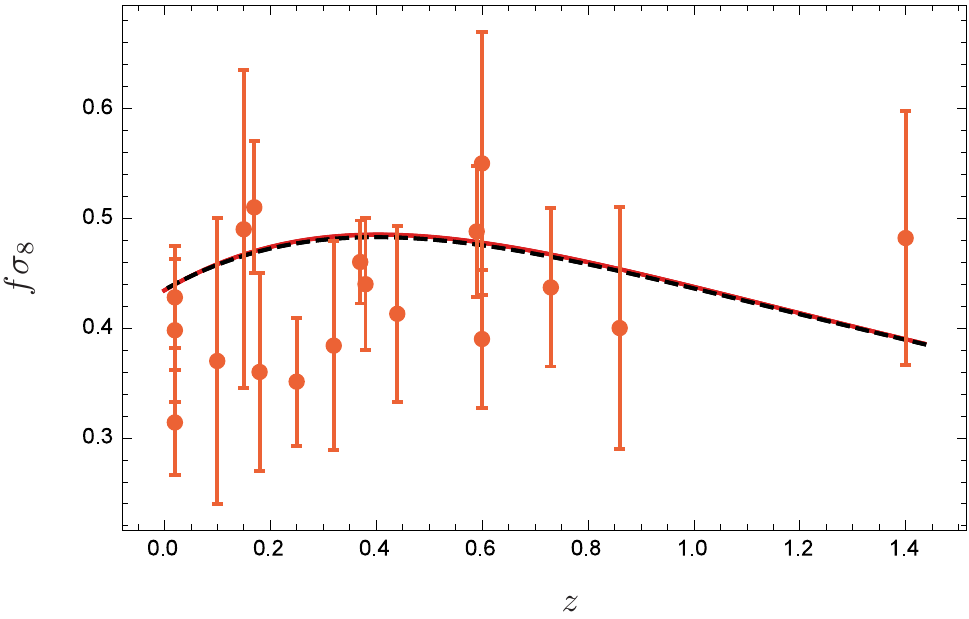}
\caption{This figure shows  the evolution of the predicted $f\sigma_{8}$ observable versus the redshift, z. The results given by  the models A, B and C are plotted in red color where the overlap is almost perfect. The result given by the $\Lambda$CDM model is plotted in black dashed line. We have included the survey data given in  table~\ref{dataset1} with the corresponding error. }\label{errorfs8}
\end{figure}

Figure~\ref{errorfs8} shows the evolution of $f\sigma_{8}$. The $\Lambda$CDM model is plotted in black dashed line while  models A, B and C are plotted in red.


%
  \begin{table} [H]
\centering
\begin{tabularx}{.85\textwidth}{ C{0.3} C{.7} C{2.6} C{0.4} }
\toprule
	{\centering{\bm{$z$}}}
	& \bm{$f\sigma_8$}
	& {\bf Survey}
	& {\bf Ref.}
\\\midrule
	$0.02\phantom{0}$
	& $0.314 \phantom{}\pm0.048\phantom{}$
	& 2MASS
	& \cite{Davis:2010sw}
 \\\midrule
	$0.02\phantom{0}$
	& $0.398 \phantom{}\pm0.065\phantom{}$
	& SNIa + IRAS
	& \cite{Turnbull:2011ty}
 \\\midrule
	$0.02\phantom{0}$
	& $0.428 \phantom{}\pm0.046\phantom{}$
	& 6dF Galaxy Survey + SNIa
	& \cite{Huterer:2016uyq}
 \\\midrule
	$0.10\phantom{0}$
	& $0.370 \phantom{}\pm0.130\phantom{}$
	& SDSS-veloc
	& \cite{Feix:2015dla}
 \\\midrule
	$0.15\phantom{0}$
	& $0.49\phantom{0}\pm0.15\phantom{0}$
	& SDSS DR7 MGS
	& \cite{Howlett:2014opa}
 \\\midrule
	$0.17\phantom{0}$
	& $0.51\phantom{0}\pm0.06\phantom{0}$
	& 2dF Galaxy Redshift Survey
	& \cite{Percival:2004fs,Song:2008qt}
 \\\midrule
	$0.18\phantom{0}$
	& $0.360\phantom{}\pm0.090\phantom{}$
	& GAMA
	& \cite{Blake:2013nif}
 \\\midrule
	$0.25\phantom{0}$
	& $0.351\pm0.058$
	& SDSS II LRG
	& \cite{Samushia:2011cs}
 \\\midrule
	$0.32\phantom{0}$
	& $0.384\pm0.095$
	&  BOSS LOWZ
	& \cite{Sanchez:2013tga}
 \\\midrule
	$0.37\phantom{0}$
	& $0.460\pm0.038$
	& SDSS II LRG
	& \cite{Samushia:2011cs}
 \\\midrule
	$0.38\phantom{0}$
	& $0.440\pm0.060$
	& GAMA
	& \cite{Blake:2013nif}
 \\\midrule
	$0.44\phantom{0}$
	& $0.413\pm0.080$
	& WiggleZ Dark Energy Survey + Alcock-Paczynski distortion
	& \cite{Blake:2012pj}
 \\\midrule
	$0.59\phantom{0}$
	& $0.488\pm0.06\phantom{0}$
	& SDSS III BOSS DR12 CMASS
	& \cite{Chuang:2013wga}
 \\\midrule
	$0.60\phantom{0}$
	& $0.390\pm0.063$
	& WiggleZ Dark Energy Survey + Alcock-Paczynski distortion
	& \cite{Blake:2012pj}
 \\\midrule
	$0.60\phantom{0}$
	& $0.550\pm0.120$
	& Vipers PDR-2
	& \cite{Pezzotta:2016gbo}
 \\\midrule
	$0.73\phantom{0}$
	& $0.437\pm0.072$
	& WiggleZ Dark Energy Survey + Alcock-Paczynski distortion
	& \cite{Blake:2012pj}
 \\\midrule
	$0.86\phantom{0}$
	& $0.400\pm0.110\phantom{}$
	& Vipers PDR-2
	& \cite{Pezzotta:2016gbo}
 \\\midrule
	$1.40\phantom{0}$
	& $0.482\pm0.116$
	& FastSound
	& \cite{Okada:2015vfa}
 \\
\bottomrule
\end{tabularx}
\caption{\label{tablegrowth}%
This table shows $f\sigma_8$ measurements from independent surveys. The first column denotes the redshift while in the second column the corresponding value of $f\sigma_8$ and its error. In the third column, we show the name of survey and in the final column the reference.
}\label{dataset1}
\end{table}
\section{Conclusions}\label{conclusions}

In this work we have analysed three genuine phantom models. We call those models as A, B and C, where each of them induces a particular future cosmological event known as a BR, a LR and a LSBR, respectively. These models are characterised by having a particular EoS, which can be understood as deviations from the widely accepted $\Lambda$CDM model and therefore, suitable models to describe the current Universe. We use SNIa, CMB, BAO and $H(z)$ data and the Markov Chain Monte Carlo method to estimate the cosmological parameters of models A, B and C. We remark that the model C  has not been constrained previously. In the case of the model A, the corresponding model parameter consists on the EoS parameter of dark energy. This value has been observationally constrained, for example,  in \cite{Ade:2015xua,wikiesa}, where the result given, $w_{\textrm{d}}=-1.019$, is very close to the one we have found, $w_{\textrm{d}}=-1.02758$. Similarly, the result obtained for the model B is of the same order of  magnitude to that found in \cite{Frampton:2011sp} where the relative difference  is less than a $6\%$,  this can be understood as an indicative of the reliability of the obtained results. In addition, we have computed the results of the $\Lambda$CDM model as well, in order to make a comparison with the models A, B and C. 

Once we have fitted observationally the parameters of the models, we have computed numerically linear cosmological  perturbations since the radiation dominated epoch. Therefore, we have not only considered perturbations of DM and DE but also those of radiation. The numerical calculations have been performed till the far future. In particular, we have obtained the density contrast of DM and DE and evaluated as well the matter power spectrum of DM and the corresponding evolution of $f\sigma_{8}$. We have as well computed the evolution of the Bardeen gravitational potential. We have confirmed that indeed in phantom DE models, it is expected that the Bardeen gravitational potential will flip its sign in the future  \cite{Albarran:2016mdu,Albarran:2017kzf}. This is in accordance with the fact that all the structures will be destroyed in phantom DE models.

Using the values of the best fit (shown in the third column of table \ref{tab1}), the  matter power spectrum  given by  models A, B and C are so similar that it is almost impossible to distinguish  them. Something similar happens when comparing the results of  $f\sigma_{8}$. In order to give an account of  small differences, we compute the relative deviation with respect to $\Lambda$CDM and found that the largest differences are around a $2\%$ for the matter power spectrum and $f\sigma_{8}$. The models  show a very similar phenomenology till the present time and future cosmic times. We notice that there are no significant differences that could allow us to find a characteristic footprint of each model  with enough accuracy.
In view of this, we compare the predicted $f\sigma_{8}$ evolution  against the observational data and compute the reduced $\chi^{2}$. We conclude that the $\Lambda$CDM model gives the best fit. The models A, B and C have similar behaviour with respect to the $\Lambda$CDM. When performing the $\chi^{2}$ analysis at $f\sigma_{8}$ level, what we found is that the model C is the one which less deviates from  $\Lambda$CDM while the model A is the one with larger deviations. However, this is not  enough to conclude that the model C is observationally preferred. We conclude as well that there is a disagreement between the background and the perturbation analysis. This discordance will be a subject of a future work.


\section{acknowledgments}

MBL is supported by the Basque Foundation of Science Ikerbasque. She also would like to acknowledge the partial support from the Basque government Grant No. IT956-16 (Spain) and from the project FIS2017-85076-P (MINECO/AEI/FEDER, UE).


\appendix
\section{Review on linear Cosmological perturbations}\label{seclinearpert}

In this section we present a brief summary of the theory of  linear cosmological perturbations. We just  consider scalar perturbations, therefore, the perturbed metric reads
\begin{equation}\label{lieneelement2}
ds^{2}=a^{2}\left\{-\left(1+2\Phi\right)d\eta^{2}+\left[\left(1-2\Psi\right)\delta_{ij}\right]dx^{i}dx^{j}\right\},
\end{equation}
where $\eta$ is the conformal time, i.e.  $d\eta=dt/a$. The quantities $\Phi$ and $\Psi$ are the Bardeen potentials. The perturbed Einstein equation reads
\begin{align}
\delta R^{\mu}_{\phantom{z}\nu}-\frac{1}{2}\delta^{\mu}_{\phantom{z}\nu}\delta R= & \ 8\pi G\delta T^{\mu}_{\phantom{z}\nu},\label{Einstein}
\end{align}
where the perturbed total energy momentum tensor, $\delta T^{\mu}_{\phantom{z}\nu}$, is the sum of the perturbed energy momentum tensors  given by the different components, i.e. $\delta T^{\mu}_{\phantom{z}\nu}=\delta T^{\mu}_{\textrm{r} \nu}+\delta T^{\mu}_{\textrm{m}\nu}+\delta T^{\mu}_{\textrm{d}\nu}$, where the subscripts  r, m, d correspond to radiation, matter and DE, respectively. Then, for a given component $\ell$, the perturbed energy momentum tensor can be written as  \cite{AmendolaTsujikawa}
\begin{align}
\delta T_{\phantom{z}\ell 0}^{0}=&-\delta \rho_{\ell}
\,\label{deltaT00},\\
\delta T_{\phantom{z}\ell  0}^{i}=&-\left(p_{\ell}+\rho_{\ell}\right)\partial^{i}v_{\ell}
\,\label{deltaT0i}, \\
\delta T_{\phantom{z}\ell  i}^{0}=&\left(p_{\ell}+\rho_{\ell}\right)\partial_{i}v_{\ell}
\,\label{deltaTii}, \\
\delta T_{\phantom{z}\ell  j}^{i}=& \ \delta p_{\ell} \ \delta_{j}^{i}+\Pi_{\,j}^{i}
\,\label{deltaTij}, 
\end{align}
where $\delta\rho_{\ell}$ is the perturbed energy density,  $p_{\ell}$ is the pressure perturbation, $\Pi_{\,j}^{i}$ is the anisotropic stress tensor and  $v_{\ell}$ is the peculiar velocity.  We will neglect the anisotropic stress tensor from now on. This leads to the equality between Bardeen potentials through Eq. (\ref{Einstein}) \cite{AmendolaTsujikawa,Malik:2008im,Bassett:2005xm}. So from now on, we will set $\Psi=\Phi$. The perturbed Einstein equations imply as well
\begin{align}
&\Psi=\frac{3\mathcal{H}^2}{2k^2}\left[3\left(1+w\right)\mathcal{H}v-\delta\right]\label{Psifinal1}\\
&\Psi_{x}=-\frac{3\mathcal{H}^2}{2k^2}\left[3\left(1+w\right)\mathcal{H}v-\delta\right]-\frac{3}{2}\left(1+w\right)\mathcal{H}v,\label{Psifinal2}
\end{align}
%
%
%
where the subscript $x$ stands for derivatives with respect to $x=\ln\left(a\right)$ and $k$ is the wave number. The total quantities as the total  EoS parameter and the total velocity field; noted without subscript, are given by the sum of the partial contributions of each component as
\begin{equation}\label{pecvelocity}
v=\sum_{\ell}\frac{1+w_{\ell}}{1+w}\Omega_{\ell}v_{\ell},\qquad\textrm{where}\qquad w=\sum_{\ell}\Omega_{\ell}w_{\ell}.
\end{equation}

On the other hand,  it is well known that a negative equation of state parameter (which is the case of  DE energy) induces classical instabilities. To avoid such instabilities we decompose the pressure  in its adiabatic and non-adiabatic contributions as \cite{Bean:2003fb,Valiviita:2008iv}
\begin{eqnarray}
\delta p_{\textrm{d}}\equiv\left[c_{s\textrm{d}}^2-\mathcal{H}\left(1+w_{\textrm{d}}\right)\left(c_{a\textrm{d}}^2-c_{s\textrm{d}}^2\right)\frac{v_{\textrm{d}}}{\delta_{\textrm{d}}}\right]\delta\rho_{\textrm{d}},\label{csgauge}
\end{eqnarray} 
where $c_{s\textrm{d}}$ is the speed of sound  in the rest frame and  $c_{a\textrm{d}}$ is the adiabatic speed of sound. These quantities are defined as 
\begin{equation}\label{defcs}
\left.c_{s\textrm{d}}^2\equiv\frac{\delta p_{\textrm{d}}}{\delta\rho_{\textrm{d}}}\right|_{\textrm{r.f}}, \qquad c_{a\textrm{d}}^2\equiv\frac{p'_{\textrm{d}}}{\rho'_{\textrm{d}}}.
\end{equation}
As can be seen, both expressions coincide in the case of a barotropic fluid description with a constant EoS parameter. As it is common in other works, we  set  $c_{s\textrm{d}}^{2}=1$.

Finally, we consider that each component is conserved separately. This leads  to a couple of equations for each cosmic component. Consequently, we get the following six first order linear differential equations
\begin{align}
	\left(\delta_{\textrm{r}}\right)_x &=\frac{4}{3}\left(\frac{k^2}{\mathcal{H}}v_{\textrm{r}}+3\Psi_x\right)
	\,\label{deltarx},\\
	\left(v_{\textrm{r}}\right)_x &=-\frac{1}{\mathcal{H}}\left(\frac{1}{4}\delta_{\textrm{r}}+\Psi\right)
	\,\label{vrx},\\
	\left(\delta_{\textrm{m}}\right)_x &=\left(\frac{k^2}{\mathcal{H}}v_{\textrm{r}}+3\Psi_x\right)
	\,\label{deltamx},\\
	\left(v_{\textrm{m}}\right)_x &=-\left(v_{\textrm{m}}+\frac{\Psi}{\mathcal{H}}\right)
	\,\label{vmx},\\
	\left(\delta_{\textrm{d}}\right)_x &=\left(1+w_{\textrm{d}}\right)\left\{\left[\frac{k^2}{\mathcal{H}} + 9\mathcal{H}\left(c_{s\textrm{d}}^2-c_{a\textrm{d}}^2\right) \right]v_{\textrm{d}}+3\Psi_x\right\}+3\left(w_{\textrm{d}}-c_{s\textrm{d}}^2\right)\delta_{\textrm{d}}
	\,\label{deltadx},\\
	\left(v_{\textrm{d}}\right)_x &=-\frac{1}{\mathcal{H}} \left(\frac{c_{s\textrm{d}}^2}{1+w_{\textrm{d}}}\delta_{\textrm{d}}+\Psi\right) +\left(3c_{s\textrm{d}}^2-1\right)v_{\textrm{d}},\label{vdx}\\ 
	 \nonumber
\end{align}
All matter perturbations  are connected  through the quantities $\Psi$ and $\Psi_x$. The initial conditions will be fixed  as done in \cite{Albarran:2016mdu}; i.e. (i) we will assume adiabatic conditions for densities contrast and velocities (cf Eqs. (3.20) and (3.21) of Ref. \cite{Albarran:2016mdu}), 
\begin{align}
	\label{adiab}
	\frac{\delta_{\textrm{r}}}{1+w_{\textrm{r}}}=\frac{\delta_{\textrm{m}}}{1+w_{\textrm{m}}}=\frac{\delta_{\textrm{d}}}{1+w_{\textrm{d}}} = \frac{\delta}{1+w}
	\,.
\end{align}
The previous equations lead to write the  initial partial density contrasts in terms of initial total density contrast, $\delta_{\textrm{i}}$, as
\begin{align}
	\label{initcond1}
	\frac{3}{4}\delta_{\textrm{r,i}}
	=
	\delta_{\textrm{m,i}}
	=
	\frac{\delta_{\textrm{d,i}}}{1+w_{\textrm{d,i}}}\approx\frac{3}{4}\delta_{\textrm{i}}
	\,,
\end{align}
(ii) these conditions will be applied for modes well inside the radiation dominated epoch and outside the horizon and (iii) the initial total matter density contrast is fixed through Planck data for a single field inflation \cite{Ade:2015xua}
\begin{equation}\label{deltaphys}
\delta_{\textrm{phys}}=\frac{4\pi}{3}\sqrt{2A_{\textrm{s}}}\left[\frac{k}{k_{\textrm{pivot}}}\right]^{\frac{n_{\textrm{s}}-1}{2}}k^{-\frac{3}{2}},
\end{equation}
where $A_{\textrm{s}}$ and $n_{\textrm{s}}$ are the amplitude  and the spectral index, respectively. We assume $A_{\textrm{s}}=2.143\times10^{-9}$ and $n_{\textrm{s}}=0.9681$ as given in Ref. \cite{Ade:2015rim}. In addition, the pivot scale is $k_{\textrm{pivot}}=0.05 \ \textrm{Mpc}^{-1}$. Finally, we remind that the matter power spectrum is defined as \cite{Wands:2009ex,Bruni:2011ta}
\begin{equation}
P_{\hat{\delta}_{\textrm{m}}}=\left\vert\delta_{\textrm{m}} - 3\mathcal{H}v_{\textrm{m}}\right\vert^2,
\end{equation}
and $f\sigma_{8}$ as
\begin{equation}\label{sigma8}
	f\sigma_8 =  \frac{\sigma_8\left(0,\,k_{\sigma_8}\right) }{\delta_{\textrm{m}}\left(0,\,k_{\sigma_8}\right)}\frac{d \delta_{\textrm{m}}\left(z,k_{\sigma_8}\right)}{d x},
\end{equation}
where $f\equiv d \left(\ln\delta_{\textrm{m}}\right)/d \left(\ln a\right)$ and $\sigma_8\left(z, k_{\sigma_8}\right)$ is \cite{Wang:2010gq}
\begin{equation}\label{sigma8}
	\sigma_8\left(z, k_{\sigma_8}\right) = \sigma_8\left(0,\,k_{\sigma_8}\right) \frac{\delta_{\textrm{m}}\left(z,k_{\sigma_8}\right) }{\delta_{\textrm{m}}\left(0,\,k_{\sigma_8}\right)}.
\end{equation}
The wave number $k_{\sigma_8}$ is set to $k_{\sigma_8}=0.125$ h Mpc$^{-1}$. In addition,  $\sigma_8\left(0, k_{\sigma_8}\right)$ corresponds to the current value of $\sigma_8$. We have set for all  models and  parameters $\sigma_8\left(0, k_{\sigma_8}\right)=0.820$, in accordance with Ref. \cite{Ade:2015xua}.


\begin{thebibliography}{}



\bibitem{Riess:1998cb}
 A.~G.~Riess {\it et al.} [Supernova Search Team Collaboration],
 Astron.\ J.\ {\bf 116} (1998) 1009
 [\href{https://arxiv.org/abs/astro-ph/9805201}{astro-ph/9805201}].
 
\bibitem{Perlmutter:1998np}
 S.~Perlmutter {\it et al.} [Supernova Cosmology Project Collaboration],
 Astrophys.\ J.\ {\bf 517} (1999) 565
 [\href{https://arxiv.org/abs/astro-ph/9812133}{astro-ph/9812133}].
 
\bibitem{Ade:2015rim}
P.~A.~R.~Ade {\it et al.} [Planck Collaboration],
  Astron.\ Astrophys.\  {\bf 594} (2016) A14
 [\href{https://arxiv.org/abs/1502.01590}{arXiv:1502.01590 [astro-ph.CO]}].
 
\bibitem{Ade:2015xua}
P.~A.~R.~Ade {\it et al.} [Planck Collaboration],
  Astron.\ Astrophys.\  {\bf 594} (2016) A13
 [\href{https://arxiv.org/abs/1502.01589}{arXiv:1502.01589 [astro-ph.CO]}].
 
\bibitem{Aghanim:2018eyx}
  N.~Aghanim {\it et al.} [Planck Collaboration],
 [\href{https://arxiv.org/abs/1807.06209}{arXiv:1807.06209 [astro-ph.CO]}].
 
\bibitem{Tsujikawa:2010sc}
 S.~Tsujikawa,
 [\href{https://arxiv.org/abs/1004.1493}{arXiv:1004.1493 [astro-ph.CO]}].
 
  \bibitem{AmendolaTsujikawa}
 L. Amendola and S. Tsujikawa,
 \textit{Dark Energy: Theory and Observations}. First edition (Cambridge University Press, Cambridge, 2010). 
 
\bibitem{Stefancic:2003rc}
  H.~\^{S}tefan\^{c}i\'{c},
  Phys.\ Lett.\ B {\bf 586} (2004) 5
 [\href{ https://arxiv.org/abs/astro-ph/0310904}{arXiv:0310904[astro-ph.CO]}].
 
\bibitem{Dabrowski:2003jm}
 M.~P.~D\c{a}browski, T.~Stachowiak, and M.~Szyd{\l }owski,
 Phys.\ Rev.\ D {\bf 68} (2003) 103519
 [\href{https://arxiv.org/abs/hep-th/0307128}{hep-th/0307128}].

\bibitem{Elizalde:2008yf}
  E.~Elizalde, S.~Nojiri, S.~D.~Odintsov, D.~Saez-Gomez and V.~Faraoni,
  Phys.\ Rev.\ D {\bf 77} (2008) 106005
  [\href{https://arxiv.org/pdf/0803.1311.pdf}{arXiv:0803.1311 [hep-th]}]. 
  
\bibitem{Caldwell:1999ew}
 R.~R.~Caldwell,
 Phys.\ Lett.\ B {\bf 545} (2002) 23
 [\href{http://arxiv.org/abs/astro-ph/9908168}{astro-ph/9908168}].
 
\bibitem{Jimenez:2016sgs}
 J.~Beltr\'an Jim\'enez, R.~Lazkoz, D.~S\'aez-G\'omez and V.~Salzano,
  Eur.\ Phys.\ J.\ C {\bf 76} (2016) no.11,  631
 [\href{https://arxiv.org/abs/1602.06211}{arXiv:1602.06211 [gr-qc]}].

 
  
\bibitem{Sahni:2014ooa}
  V.~Sahni, A.~Shafieloo and A.~A.~Starobinsky,
  Astrophys.\ J.\  {\bf 793} (2014) no.2,  L40
  [\href{https://arxiv.org/pdf/1406.2209.pdf}{arXiv:1406.2209 [astro-ph.CO]}].
  
  
   
 
\bibitem{Vagnozzi:2018jhn}
  S.~Vagnozzi, S.~Dhawan, M.~Gerbino, K.~Freese, A.~Goobar and O.~Mena,
  Phys.\ Rev.\ D {\bf 98} (2018) no.8,  083501
  [\href{https://arxiv.org/pdf/1801.08553.pdf}{arXiv:1801.08553 [astro-ph.CO]}].
  
\bibitem{Alam:2016wpf}
  U.~Alam, S.~Bag and V.~Sahni,
  Phys.\ Rev.\ D {\bf 95} (2017) no.2,  023524
  [\href{https://arxiv.org/pdf/1605.04707.pdf}{arXiv:1605.04707 [astro-ph.CO]}].
 

 


\bibitem{Starobinsky:1999yw}
 A.~A.~Starobinsky,
 Grav.\ Cosmol.\ {\bf 6} (2000) 157
 [\href{http://arxiv.org/abs/astro-ph/9912054}{astro-ph/9912054}].
 
\bibitem{Caldwell:2003vq}
 R.~R.~Caldwell, M.~Kamionkowski and N.~N.~Weinberg,
 Phys.\ Rev.\ Lett.\ {\bf 91} (2003) 071301
 [\href{https://arxiv.org/abs/astro-ph/0302506}{astro-ph/0302506}].

\bibitem{Carroll:2003st}
 S.~M.~Carroll, M.~Hoffman and M.~Trodden,
 Phys.\ Rev.\ D {\bf 68} (2003) 023509
 [\href{http://arxiv.org/abs/astro-ph/0301273}{astro-ph/0301273}].
 
\bibitem{Chimento:2003qy}
 L.~P.~Chimento and R.~Lazkoz,
 Phys.\ Rev.\ Lett.\ {\bf 91} (2003) 211301
 [\href{http://arxiv.org/abs/gr-qc/0307111}{gr-qc/0307111}].

\bibitem{GonzalezDiaz:2003rf}
 P.~F.~Gonz\'{a}lez-D\'{i}az,
 Phys.\ Lett.\ B {\bf 586} (2004) 1
 [\href{http://arxiv.org/abs/astro-ph/0312579}{astro-ph/0312579}].

\bibitem{GonzalezDiaz:2004vq}
 P.~F.~Gonz\'{a}lez-D\'{i}az,
 Phys.\ Rev.\ D {\bf 69} (2004) 063522
 [\href{https://arxiv.org/abs/hep-th/0401082}{hep-th/0401082}].
 
\bibitem{Sahni:2002dx}
  V.~Sahni and Y.~Shtanov,
  JCAP {\bf 0311} (2003) 014
 [\href{https://arxiv.org/pdf/astro-ph/0202346.pdf}{[astro-ph/0202346]}].
 
 \bibitem{Ruzmaikina}
T. Ruzmaikina and A. A. Ruzmaikin, Sov. Phys. JETP {\bf 30} (1970) 372.
 
 
\bibitem{Bouhmadi-Lopez:2013nma}
 M.~Bouhmadi-L\'{o}pez, P.~Chen, and Y.~W.~Liu,
 Eur.\ Phys.\ J.\ C {\bf 73} (2013) 2546
 [\href{https://arxiv.org/abs/1302.6249}{arXiv:1302.6249 [gr-qc]}].

 
\bibitem{Nojiri:2005sx}
 S.~'i.~Nojiri, S.~D.~Odintsov and S.~Tsujikawa,
 Phys.\ Rev.\ D {\bf 71} (2005) 063004
 [\href{https://arxiv.org/abs/hep-th/0501025}{hep-th/0501025}].
  
\bibitem{Nojiri:2005sr}
 S.~'i.~Nojiri and S.~D.~Odintsov,
 Phys.\ Rev.\ D {\bf 72} (2005) 023003
 [\href{https://arxiv.org/abs/hep-th/0505215}{hep-th/0505215}].
 
\bibitem{Stefancic:2004kb}
 H.~\v{S}tefan\v{c}i{\'c},
 Phys.\ Rev.\ D {\bf 71} (2005) 084024
 [\href{https://arxiv.org/abs/astro-ph/0411630}{astro-ph/0411630}].
 
\bibitem{BouhmadiLopez:2005gk}
 M.~Bouhmadi-L{\'o}pez,
 Nucl.\ Phys.\ B {\bf 797} (2008) 78
 [\href{https://arxiv.org/abs/astro-ph/0512124}{astro-ph/0512124}].
 
\bibitem{Frampton:2011sp}
 P.~H.~Frampton, K.~J.~Ludwick and R.~J.~Scherrer,
 Phys.\ Rev.\ D {\bf 84} (2011) 063003
 [\href{https://arxiv.org/abs/1106.4996}{arXiv:1106.4996 [astro-ph.CO]}].
 

\bibitem{Brevik:2011mm}
 I.~Brevik, E.~Elizalde, S.~'i.~Nojiri, and S.~D.~Odintsov,
 Phys.\ Rev.\ D {\bf 84} (2011) 103508
 [\href{https://arxiv.org/abs/1107.4642}{arXiv:1107.4642 [hep-th]}].
 
\bibitem{Contreras:2018two}
F.~Contreras, N.~Cruz, E.~Elizalde, E.~Gonz\'alez and S.~Odintsov,
  Phys.\ Rev.\ D {\bf 98} (2018) no.12,  123520
 [\href{https://arxiv.org/abs/1808.06546}{arXiv:1808.06546 [gr-qc]}].
 
\bibitem{Bouhmadi-Lopez:2014cca}
 M.~Bouhmadi-L\'opez, A.~Errahmani, P.~Mart\'{i}n-Moruno, T.~Ouali, and Y.~Tavakoli,
 Int.\ J.\ Mod.\ Phys.\ D {\bf 24} (2015) no.10, 1550078
 [\href{https://arxiv.org/abs/1407.2446}{arXiv:1407.2446 [gr-qc]}].

 
 
\bibitem{Morais:2016bev}
  J.~Morais, M.~Bouhmadi-L\'opez, K.~Sravan Kumar, J.~Marto and Y.~Tavakoli,
  Phys.\ Dark Univ.\  {\bf 15} (2017) 7
 [\href{http://arxiv.org/abs/1608.01679}{arXiv:1608.01679 [gr-qc]}].
 
\bibitem{Bouhmadi-Lopez:2018lly}
  M.~Bouhmadi-L\'{o}pez, D.~Brizuela and I.~Garay,
  JCAP {\bf 1809} (2018) no.09,  031
  [\href{https://arxiv.org/abs/1802.05164}{ [arXiv:1802.05164 [gr-qc]]}].
 
 
\bibitem{Dabrowski:2006dd}
 M.~P.~D\c{a}browski, C.~Kiefer and B.~Sandh\"ofer,
  Phys.\ Rev.\ D {\bf 74} (2006) 044022
 [\href{https://arxiv.org/abs/hep-th/0605229}{arXiv: hep-th/0605229}].
 

\bibitem{Kamenshchik:2007zj}
 A.~Kamenshchik, C.~Kiefer and B.~Sandh\"ofer,
 Phys.\ Rev.\ D {\bf 76} (2007) 064032
 [\href{https://arxiv.org/abs/0705.1688}{arXiv:0705.1688 [gr-qc]}].

\bibitem{BouhmadiLopez:2009pu}
 M.~Bouhmadi-L\'opez, C.~Kiefer, B.~Sandh\"ofer and P.~Vargas Moniz,
 Phys.\ Rev.\ D {\bf 79} (2009) 124035
 [\href{https://arxiv.org/abs/0905.2421}{arXiv:0905.2421 [gr-qc]}].
 
\bibitem{Kamenshchik:2012ij}
 A.~Y.~Kamenshchik and S.~Manti,
 Phys.\ Rev.\ D {\bf 85} (2012) 123518
 [\href{https://arxiv.org/abs/1202.0174}{arXiv:1202.0174 [gr-qc]}].
 
\bibitem{Kamenshchik:2013naa}
 A.~Y.~Kamenshchik,
 Class.\ Quant.\ Grav.\ {\bf 30} (2013) 173001
 [\href{https://arxiv.org/abs/1307.5623}{arXiv:1307.5623 [gr-qc]}].
 
 
\bibitem{Bouhmadi-Lopez:2013tua}
 M.~Bouhmadi-L\'opez, C.~Kiefer and M.~Kr\"amer,
 Phys.\ Rev.\ D {\bf 89} (2014) no.6, 064016
 [\href{https://arxiv.org/abs/1312.5976}{arXiv:1312.5976 [gr-qc]}].
 
\bibitem{Albarran:2015tga}
 I.~Albarran and M.~Bouhmadi-L\'opez,
 JCAP {\bf 1508} (2015) no.08, 051
 [\href{https://arxiv.org/abs/1505.01353}{arXiv:1505.01353 [gr-qc]}].
 
\bibitem{Albarran:2015cda}
 I.~Albarran, M.~Bouhmadi-L\'opez, F.~Cabral and P.~Mart\'in-Moruno,
 JCAP {\bf 1511} (2015) no.11, 044
 [\href{https://arxiv.org/abs/1509.07398}{arXiv:1509.07398 [gr-qc]}].

 
\bibitem{Albarran:2016ewi}
 I.~Albarran, M.~Bouhmadi-L\'opez, C.~Kiefer, J.~Marto and P.~Vargas Moniz,
 Phys.\ Rev.\ D {\bf 94} (2016) 063536
[\href{https://arxiv.org/abs/1604.08365}{arXiv:1604.08365 [gr-qc]}].


\bibitem{Bouhmadi-Lopez:2016dcf}
 M.~Bouhmadi-L\'opez and C.~Y.~Chen,
 JCAP {\bf 1611} (2016) no.11, 023
 [\href{https://arxiv.org/abs/1609.00700}{arXiv:1609.00700 [gr-qc]}].
 
\bibitem{Albarran:2017swy}
  I.~Albarran, M.~Bouhmadi-L\'opez, C.~Y.~Chen and P.~Chen,
  Phys.\ Lett.\ B {\bf 772} (2017) 814
 [\href{https://arxiv.org/abs/1703.09263}{arXiv:1703.09263 [gr-qc]}].
 
\bibitem{Alonso-Serrano:2018zpi}
  A.~Alonso-Serrano, M.~Bouhmadi-L\'opez and P.~Mart\'in-Moruno,
  Phys.\ Rev.\ D {\bf 98} (2018) no.10,  104004
 [\href{https://arxiv.org/abs/1802.03290}{arXiv:1802.03290 [gr-qc]}].
 
\bibitem{Bouhmadi-Lopez:2018tel}
  M.~Bouhmadi-L\'opez, C.~Y.~Chen and P.~Chen,
  JCAP {\bf 1812} (2018) no.12,  032
 [\href{https://arxiv.org/abs/1810.10918}{arXiv:1810.10918 [gr-qc]}].
 
\bibitem{Albarran:2018mpg}
  I.~Albarran, M.~Bouhmadi-L\'opez, C.~Y.~Chen and P.~Chen,
  Phys.\ Dark Univ.\  100255
   [\href{https://arxiv.org/abs/1811.05041}{arXiv:1811.05041 [gr-qc]}].
   
\bibitem{Elizalde:2004mq}
  E.~Elizalde, S.~Nojiri and S.~D.~Odintsov,
  Phys.\ Rev.\ D {\bf 70} (2004) 043539
  [\href{https://arxiv.org/pdf/hep-th/0405034.pdf}{hep-th/0405034}].
  
\bibitem{Elizalde:2005ju}
  E.~Elizalde, S.~Nojiri, S.~D.~Odintsov and P.~Wang,
  Phys.\ Rev.\ D {\bf 71} (2005) 103504
  [\href{https://arxiv.org/pdf/hep-th/0502082.pdf}{hep-th/0502082}].
   
\bibitem{Bouhmadi-Lopez:2019zvz}
  M.~Bouhmadi-L\'opez, C.~Kiefer and P.~Mart\'in-Moruno,
  arXiv:1904.01836 [gr-qc].
   [\href{https://arxiv.org/abs/1904.01836}{arXiv:1904.01836 [gr-qc]}].
 
  
 \bibitem{wikiesa}
Planck 2015 Results: Cosmological Parameter Tables:\\
 \href{https://wiki.cosmos.esa.int/planckpla2015/images/0/07/Params_table_2015_limit95.pdf}{wiki.cosmos.esa.int/planckpla2015/images/0/07/Params\_table\_2015\_limit95.pdf}
 %
 
\bibitem{Balcerzak:2012ae}
 A.~Balcerzak and T.~Denkiewicz,
 Phys.\ Rev.\ D {\bf 86} (2012) 023522
 [\href{http://arxiv.org/abs/1202.3280}{arXiv:1202.3280 [astro-ph.CO]}].
 
\bibitem{Denkiewicz:2014aca}
 T.~Denkiewicz,
 JCAP {\bf 1503} (2015) 037
 [\href{https://arxiv.org/abs/1411.6169}{arXiv:1411.6169 [astro-ph.CO]}].
 
\bibitem{Denkiewicz:2015nai}
 T.~Denkiewicz,
 [\href{https://arxiv.org/abs/1511.04708}{arXiv:1511.04708 [astro-ph.CO]}].
 
\bibitem{Denkiewicz:2017ixd}
  T.~Denkiewicz and V.~Salzano,
  arXiv:1702.01291 [astro-ph.CO].
 [\href{https://arxiv.org/abs/1702.01291}{arXiv:1702.01291 [astro-ph.CO]}].
 
\bibitem{Astashenok:2012iy}
 A.~V.~Astashenok and S.~D.~Odintsov,
 Phys.\ Lett.\ B {\bf 718} (2013) 1194
 [\href{https://arxiv.org/abs/1211.1888}{arXiv:1211.1888 [gr-qc]}].
 
\bibitem{Kunz:2006wc}
 M.~Kunz and D.~Sapone,
 Phys.\ Rev.\ D {\bf 74} (2006) 123503
 [\href{https://arxiv.org/abs/astro-ph/0609040}{astro-ph/0609040}].
 
\bibitem{Bean:2003fb}
 R.~Bean and O.~Dor\'{e},
 Phys.\ Rev.\ D {\bf 69} (2004) 083503
 [\href{https://arxiv.org/abs/astro-ph/0307100}{astro-ph/0307100}].
 
\bibitem{Valiviita:2008iv}
 J.~V\"{a}liviita, E.~Majerotto and R.~Maartens,
 JCAP {\bf 0807} (2008) 020
 [\href{https://arxiv.org/abs/0804.0232}{arXiv:0804.0232 [astro-ph]}].
 
\bibitem{Albarran:2016mdu}
  I.~Albarran, M.~Bouhmadi-L\'{o}pez and J.~Morais,
  Phys.\ Dark Univ.\  {\bf 16} (2017) 94
  [\href{https://arxiv.org/abs/arXiv:1611.00392}{arXiv:1611.00392 [astro-ph.CO]}].
  
\bibitem{Velten:2017mtr}
  H.~Velten and R.~Fazolo,
  Phys.\ Rev.\ D {\bf 96} (2017) no.8,  083502
 [\href{https://arxiv.org/abs/1707.03224}{ [arXiv:1707.03224 [astro-ph.CO]]}].
 
\bibitem{Maeder:2018xww}
  A.~Maeder and V.~G.~Gueorguiev,
 [\href{https://arxiv.org/abs/1811.03495}{ [arXiv:1811.03495 [astro-ph.CO]}].
 
\bibitem{Mifsud:2017fsy}
  J.~Mifsud and C.~Van De Bruck,
  JCAP {\bf 1711} (2017) no.11,  001
[\href{https://arxiv.org/abs/1707.07667}{[arXiv:1707.07667 [astro-ph.CO]]}].
 
\bibitem{Dutta:2017wfd}
  J.~Dutta, W.~Khyllep and N.~Tamanini,
  JCAP {\bf 1801} (2018) 038
  [\href{https://arxiv.org/abs/1707.09246}{  [arXiv:1707.09246 [gr-qc]]}]. 
  
\bibitem{Ferreira:2018wup}
  E.~Ferreira, G.M., G.~Franzmann, J.~Khoury and R.~Brandenberger,
  [\href{https://arxiv.org/abs/1810.09474}{arXiv:1810.09474 [astro-ph.CO]}]. 
  
\bibitem{Battye:2017ysh}
  R.~A.~Battye, B.~Bolliet and F.~Pace,
  Phys.\ Rev.\ D {\bf 97} (2018) no.10,  104070
[\href{https://arxiv.org/abs/1712.05976}{  [arXiv:1712.05976 [astro-ph.CO]]}]. 




\bibitem{Morais:2015ooa}
  J.~Morais, M.~Bouhmadi-L\'opez and S.~Capozziello,
  JCAP {\bf 1509} (2015) no.09,  041
[\href{https://arxiv.org/abs/1507.02623}{arXiv:1507.02623 [gr-qc]}]. 

\bibitem{Arjona:2018jhh}
R.~Arjona, W.~Cardona and S.~Nesseris,
  Phys.\ Rev.\ D {\bf 99} (2019) no.4,  043516
[\href{https://arxiv.org/abs/1811.02469}{  arXiv:1811.02469 [astro-ph.CO]}]. 

\bibitem{delaMacorra:2018zbk}
  A.~de la Macorra and E.~Almaraz,
  Phys.\ Rev.\ Lett.\  {\bf 121} (2018) no.16,  161303
[\href{https://arxiv.org/abs/1805.01510}{  [arXiv:1805.01510 [astro-ph.CO]]}]. 

\bibitem{Scolnic:2017caz} 
   D.~M.~Scolnic {\it et al.},
  Astrophys.\ J.\  {\bf 859} (2018) no.2,  101
  [\href{https://arxiv.org/abs/1710.00845}{arXiv:1710.00845 [astro-ph.CO]}].

  \bibitem{Zhai:2018vmm} 
  Z.~Zhai and Y.~Wang,
  JCAP {\bf 1907} (2019) no.07,  005
 [\href{https://arxiv.org/abs/1811.07425v2}{arXiv:1811.07425 [astro-ph.CO]}].  
     
\bibitem{Kazantzidis:2018jtb}
  L.~Kazantzidis, L.~Perivolaropoulos and F.~Skara,
  Phys.\ Rev.\ D {\bf 99} (2019) no.6,  063537
  [\href{https://arxiv.org/pdf/1812.05356.pdf}{arXiv:1812.05356 [astro-ph.CO]}].
  
     \bibitem{Anderson:2013zyy} 
   L.~Anderson {\it et al.} [BOSS Collaboration],
  Mon.\ Not.\ Roy.\ Astron.\ Soc.\  {\bf 441} (2014) no.1,  24
 [\href{https://arxiv.org/abs/1312.4877} {arXiv:1312.4877 [astro-ph.CO]}].     
                 
                             
\bibitem{Beutler:2011hx} 
  F.~Beutler {\it et al.},
  Mon.\ Not.\ Roy.\ Astron.\ Soc.\  {\bf 416} (2011) 3017
  [\href{https://arxiv.org/abs/1106.3366}{arXiv:1106.3366 [astro-ph.CO]}].
  
       
\bibitem{Ross:2014qpa} 
  A.~J.~Ross, L.~Samushia, C.~Howlett, W.~J.~Percival, A.~Burden and M.~Manera,
  Mon.\ Not.\ Roy.\ Astron.\ Soc.\  {\bf 449} (2015) no.1,  835
  [\href{https://arxiv.org/abs/1409.3242}{arXiv:1409.3242 [astro-ph.CO]}].
 
   
\bibitem{Kazin:2014qga} 
  E.~A.~Kazin {\it et al.},
  Mon.\ Not.\ Roy.\ Astron.\ Soc.\  {\bf 441} (2014) no.4,  3524
   [\href{https://arxiv.org/abs/1401.0358}{arXiv:1401.0358 [astro-ph.CO]}].
                   

 \bibitem{Alam:2016hwk} 
 S.~Alam {\it et al.} [BOSS Collaboration],
  Mon.\ Not.\ Roy.\ Astron.\ Soc.\  {\bf 470} (2017) no.3,  2617
  [\href{https://arxiv.org/abs/1607.03155}{arXiv:1607.03155 [astro-ph.CO]}].
  
 \bibitem{Zhang:2012mp} 
  C.~Zhang, H.~Zhang, S.~Yuan, T.~J.~Zhang and Y.~C.~Sun,
  Res.\ Astron.\ Astrophys.\  {\bf 14} (2014) no.10,  1221
  [\href{https://arxiv.org/abs/1207.4541}{arXiv:1207.4541 [astro-ph.CO]}].  
  
\bibitem{Stern:2009ep} 
  D.~Stern, R.~Jimenez, L.~Verde, M.~Kamionkowski and S.~A.~Stanford,
  JCAP {\bf 1002} (2010) 008
  [\href{https://arxiv.org/abs/0907.3149}{arXiv:0907.3149 [astro-ph.CO]}].
  
\bibitem{Moresco:2012jh} 
 M.~Moresco {\it et al.},
  JCAP {\bf 1208} (2012) 006
  [\href{https://arxiv.org/abs/1201.3609}{arXiv:1201.3609 [astro-ph.CO]}].
  
\bibitem{Chuang:2012qt} 
 C.~H.~Chuang and Y.~Wang,
  Mon.\ Not.\ Roy.\ Astron.\ Soc.\  {\bf 435} (2013) 255
  [\href{https://arxiv.org/abs/1209.0210}{arXiv:1209.0210 [astro-ph.CO]}].
  
\bibitem{Moresco:2015cya} 
  M.~Moresco,
  Mon.\ Not.\ Roy.\ Astron.\ Soc.\  {\bf 450} (2015) no.1,  L16
  [\href{https://arxiv.org/abs/1503.01116}{arXiv:1503.01116 [astro-ph.CO]}].
  
   \bibitem{Moresco:2016mzx} 
  M.~Moresco {\it et al.},
  JCAP {\bf 1605} (2016) no.05,  014
  [\href{https://arxiv.org/abs/1601.01701}{arXiv:1601.01701 [astro-ph.CO]}].
  
\bibitem{Stocker:2018avm} 
    P.~Stöcker, M.~Krämer, J.~Lesgourgues and V.~Poulin,
  JCAP {\bf 1803} (2018) no.03,  018
  [\href{https://arxiv.org/abs/1801.01871}{arXiv:1801.01871 [astro-ph.CO]}].
  

  
\bibitem{Conley:2011ku} 
  A.~Conley {\it et al.} [SNLS Collaboration],
  Astrophys.\ J.\ Suppl.\  {\bf 192} (2011) 1
  [\href{https://arxiv.org/abs/1104.1443}{arXiv:1104.1443 [astro-ph.CO]}].
  
\bibitem{Komatsu:2008hk} 
  E.~Komatsu {\it et al.} [WMAP Collaboration],
  Astrophys.\ J.\ Suppl.\  {\bf 180} (2009) 330
  [\href{https://arxiv.org/abs/0803.0547}{arXiv:0803.0547 [astro-ph]}].

  
\bibitem{Fixsen:2009ug}
   D.~J.~Fixsen,
  Astrophys.\ J.\  {\bf 707} (2009) 916
   [\href{https://arxiv.org/abs/0911.1955}{arXiv:0911.1955 [astro-ph.CO]}].
   
   
\bibitem{Hu:1995en}
  W.~Hu and N.~Sugiyama,
  Astrophys.\ J.\  {\bf 471} (1996) 542
   [\href{https://arxiv.org/abs/astro-ph/9510117}{astro-ph/9510117}].
  
  

  \bibitem{Eisenstein:2005su} 
 D.~J.~Eisenstein {\it et al.} [SDSS Collaboration],
  Astrophys.\ J.\  {\bf 633} (2005) 560
  [\href{https://arxiv.org/abs/astro-ph/0501171}{astro-ph/0501171}].


\bibitem{Eisenstein:1997ik}
  D.~J.~Eisenstein and W.~Hu,
  Astrophys.\ J.\  {\bf 496} (1998) 605
  [\href{https://arxiv.org/abs/astro-ph/9709112}{astro-ph/9709112}].


  
\bibitem{Gaztanaga:2008xz} 
  E.~Gaztanaga, A.~Cabre and L.~Hui,
  Mon.\ Not.\ Roy.\ Astron.\ Soc.\  {\bf 399} (2009) 1663
  [\href{https://arxiv.org/abs/0807.3551}{arXiv:0807.3551 [astro-ph]}].
  
  
\bibitem{Jimenez:2001gg} 
  R.~Jimenez and A.~Loeb,
  Astrophys.\ J.\  {\bf 573} (2002) 37
  [\href{https://arxiv.org/abs/astro-ph/0106145}{[astro-ph/0106145]}].
  
   \bibitem{Blake:2012pj} 
 C.~Blake {\it et al.},
  Mon.\ Not.\ Roy.\ Astron.\ Soc.\  {\bf 425} (2012) 405
  [\href{https://arxiv.org/abs/1204.3674}{arXiv:1204.3674 [astro-ph.CO]}].

\bibitem{AIC} 
H. Akaike,
 [\href{https://ieeexplore.ieee.org/document/1100705}{https://ieeexplore.ieee.org/document/1100705}].

\bibitem{Basilakos:2017rgc}
  S.~Basilakos and S.~Nesseris,
  Phys.\ Rev.\ D {\bf 96} (2017) no.6,  063517
  [\href{https://arxiv.org/pdf/1705.08797.pdf}{arXiv:1705.08797 [astro-ph.CO]}].
  
 
\bibitem{Sagredo:2018rvc}
  B.~Sagredo, J.~S.~Lafaurie and D.~Sapone,
  [\href{https://arxiv.org/pdf/1808.05660.pdf}{arXiv:1808.05660 [astro-ph.CO]}].
  

\bibitem{Albarran:2017kzf}
  I.~Albarran, M.~Bouhmadi-L\'{o}pez and J.~Morais,
  Eur.\ Phys.\ J.\ C {\bf 78} (2018) no.3,  260
  [\href{https://arxiv.org/abs/1706.01484}{arXiv:1706.01484 [gr-qc]}].
  
  
\bibitem{Nesseris:2017vor}
  S.~Nesseris, G.~Pantazis and L.~Perivolaropoulos,
  Phys.\ Rev.\ D {\bf 96} (2017) no.2,  023542
  [\href{https://arxiv.org/abs/1703.10538}{arXiv:1703.10538 [astro-ph.CO]}].
  
\bibitem{Kazantzidis:2018rnb}
  L.~Kazantzidis and L.~Perivolaropoulos,
  Phys.\ Rev.\ D {\bf 97} (2018) no.10,  103503
  [\href{https://arxiv.org/pdf/1803.01337.pdf}{arXiv:1803.01337 [astro-ph.CO]}]. 
  
  
\bibitem{Davis:2010sw}
  M.~Davis, A.~Nusser, K.~Masters, C.~Springob, J.~P.~Huchra and G.~Lemson,
  Mon.\ Not.\ Roy.\ Astron.\ Soc.\  {\bf 413} (2011) 2906
  [\href{https://arxiv.org/abs/1011.3114}{arXiv:1011.3114 [astro-ph.CO]}].
  

\bibitem{Turnbull:2011ty}
  S.~J.~Turnbull, M.~J.~Hudson, H.~A.~Feldman, M.~Hicken, R.~P.~Kirshner and R.~Watkins,
  Mon.\ Not.\ Roy.\ Astron.\ Soc.\  {\bf 420} (2012) 447
  [\href{https://arxiv.org/abs/1111.0631}{arXiv:1111.0631 [astro-ph.CO]}].
  
\bibitem{Huterer:2016uyq}
  D.~Huterer, D.~Shafer, D.~Scolnic and F.~Schmidt,
  JCAP {\bf 1705} (2017) no.05,  015
  [\href{https://arxiv.org/abs/1611.09862}{arXiv:1611.09862 [astro-ph.CO]}].
  
\bibitem{Feix:2015dla}
  M.~Feix, A.~Nusser and E.~Branchini,
  Phys.\ Rev.\ Lett.\  {\bf 115} (2015) no.1,  011301
  [\href{https://arxiv.org/abs/arXiv:1503.05945}{arXiv:1503.05945 [astro-ph.CO]}].
  
\bibitem{Howlett:2014opa}
 C.~Howlett, A.~Ross, L.~Samushia, W.~Percival and M.~Manera,
 Mon.\ Not.\ Roy.\ Astron.\ Soc.\ {\bf 449} (2015) no.1, 848
 [\href{https://arxiv.org/abs/1409.3238}{arXiv:1409.3238 [astro-ph.CO]}].
 
\bibitem{Percival:2004fs}
 W.~J.~Percival {\it et al.} [2dFGRS Collaboration],
 Mon.\ Not.\ Roy.\ Astron.\ Soc.\ {\bf 353} (2004) 1201
 [\href{http://arxiv.org/abs/astro-ph/0406513}{astro-ph/0406513}].
 
 
\bibitem{Song:2008qt}
 Y.~S.~Song and W.~J.~Percival,
 JCAP {\bf 0910} (2009) 004
 [\href{https://arxiv.org/abs/0807.0810}{arXiv:0807.0810 [astro-ph]}].


 
\bibitem{Blake:2013nif}
  C.~Blake {\it et al.},
  Mon.\ Not.\ Roy.\ Astron.\ Soc.\  {\bf 436} (2013) 3089
   [\href{https://arxiv.org/abs/1309.5556}{[arXiv:1309.5556 [astro-ph.CO]]}].
  
\bibitem{Samushia:2011cs}
 L.~Samushia, W.~J.~Percival and A.~Raccanelli,
 Mon.\ Not.\ Roy.\ Astron.\ Soc.\ {\bf 420} (2012) 2102
 [\href{https://arxiv.org/abs/1102.1014}{arXiv:1102.1014 [astro-ph.CO]}].
 
\bibitem{Sanchez:2013tga}
  A.~G.~S\'{a}nchez {\it et al.},
  Mon.\ Not.\ Roy.\ Astron.\ Soc.\  {\bf 440} (2014) no.3,  2692
   [\href{https://arxiv.org/abs/1312.4854}{arXiv:1312.4854 [astro-ph.CO]}].
  

  
\bibitem{Chuang:2013wga}
 C.~H.~Chuang {\it et al.},
 Mon.\ Not.\ Roy.\ Astron.\ Soc.\ {\bf 461} (2016) no.4, 3781
 [\href{https://arxiv.org/abs/1312.4889}{arXiv:1312.4889 [astro-ph.CO]}].
 
 
 
\bibitem{Pezzotta:2016gbo}
  A.~Pezzotta {\it et al.},
  Astron.\ Astrophys.\  {\bf 604} (2017) A33
   [\href{https://arxiv.org/abs/1612.05645}{arXiv:1612.05645 [astro-ph.CO]}].
  
\bibitem{Okada:2015vfa}
 H.~Okada {\it et al.},
 Publ.\ Astron.\ Soc.\ Jap.\ {\bf 68} (2016) no.33, id.47, 17
 [\href{http://arxiv.org/abs/1504.05592}{arXiv:1504.05592 [astro-ph.GA]}].
 


\bibitem{Malik:2008im}
  K.~A.~Malik and D.~Wands,
  Phys.\ Rept.\  {\bf 475} (2009) 1
 [\href{https://arxiv.org/abs/0809.4944}{arXiv:0809.4944 [astro-ph.CO]}].
 
\bibitem{Bassett:2005xm}
  B.~A.~Bassett, S.~Tsujikawa and D.~Wands,
  Rev.\ Mod.\ Phys.\  {\bf 78} (2006) 537
 [\href{https://arxiv.org/abs/astro-ph/0507632}{arXiv:astro-ph/0507632}].
 
 
\bibitem{Wands:2009ex}
  D.~Wands and A.~Slosar,
  Phys.\ Rev.\ D {\bf 79} (2009) 123507
 [\href{https://arxiv.org/abs/0902.1084}{arXiv:0902.1084 [astro-ph.CO]}].
 
\bibitem{Bruni:2011ta}
  M.~Bruni, R.~Crittenden, K.~Koyama, R.~Maartens, C.~Pitrou and D.~Wands,
  Phys.\ Rev.\ D {\bf 85} (2012) 041301
   [\href{https://arxiv.org/abs/1106.3999}{arXiv:1106.3999 [astro-ph.CO]}].
   
      
\bibitem{Wang:2010gq}
  Y.~Wang {\it et al.},
  Mon.\ Not.\ Roy.\ Astron.\ Soc.\  {\bf 409} (2010) 737
 [\href{https://arxiv.org/abs/1006.3517}{[arXiv:1006.3517 [astro-ph.CO]}].
 
   
 
 
 
 
 

  


  

 

  

  

 

  

 

  
  



 

 
  



 
  

 



 
 

 
 
\end{thebibliography}
\end{document}